\documentclass[12pt,english]{article}

\usepackage[maccyr]{inputenc}
\usepackage{amssymb,amsmath}
\usepackage{cite}
\usepackage{color}
\usepackage{setspace}
\usepackage{hyperref}

\def\hybrid{
        \topmargin -20pt
        \oddsidemargin 0pt
        \headheight 0pt \headsep 0pt
        \textwidth 6.25in 
        \textheight 9.5in 
        \marginparwidth .875in
        \parskip 5pt plus 1pt \jot = 1.5ex}

\hybrid

\def\be{\begin{equation}}
\def\ee{\end{equation}}
\def\bea{\begin{eqnarray}}
\def\eea{\end{eqnarray}}
\def\ba{\begin{array}}
\def\ea{\end{array}}

\begin{document}
\begin{spacing}{1.0}
\begin{titlepage}
\samepage{
\setcounter{page}{1}
\vspace{-1cm}
\begin{flushright}
{\small
ROM2F/2013/07 \\
MPP-2013-182\\
LPTENS-13/16\\
LMU-ASC 46/13\\
}
\end{flushright}
\begin{center}
\vspace{0.4cm}
{\Large \bf  Gauged supergravities and non-geometric $Q/R$-fluxes\vspace{0.2cm}\\ from asymmetric orbifold CFT's}

\vspace{8mm}
\begin{minipage}{0.97\linewidth}
\begin{center}
 {\normalsize\bf Cezar Condeescu$^{1,2}$,~ Ioannis Florakis$^{3}$, Costas Kounnas$^{4}$ and Dieter L\"{u}st$^{3,5}$ \\
}

\end{center}
\vspace{.2cm} \hspace{2.0cm}
\begin{minipage}{1.\linewidth}
\begin{footnotesize}
\begin{itemize}
     \item[${}^1$]  Sezione INFN e Dipartimento di Fisica\\
       Universit\`a di Roma ``Tor Vergata"\\
          Via della Ricerca Scientifica 1, 00133 Roma, Italy
	\item[$^{2}$]   Department of Theoretical Physics \\
 ``Horia Hulubei" National Institute of Physics and Nuclear Engineering \\
 P.O. Box MG-6, M\u{a}gurele - Bucharest, 077125, Jud. Ilfov, Rom\^{a}nia    
	 \item[$^{3}$]  Max-Planck-Institut f\"ur Physik,\\ Werner-Heisenberg-Institut, 80805 M\"unchen, Germany   
	 \item[$^{4}$] Laboratoire de Physique Th\'eorique, Ecole Normale Sup\'erieure,\\ 24 rue Lhomond, F-75231 Paris Cedex 05, France
	\item[$^{5}$]  Arnold Sommerfeld Center for Theoretical Physics\\
	 Fakult\"at f\"ur Physik, Ludwig-Maximilians-Universit\"at M\"unchen \\
	 Theresienstr. 37, 80333 M\"unchen, Germany  
\end{itemize}
\end{footnotesize}
 \vspace{0.1cm}
 \end{minipage}
\end{minipage}

\vspace{-3mm}
\begin{abstract}\vspace{1mm}
{\normalsize We investigate the orbifold limits of string theory compactifications with geometric and non-geometric fluxes. Exploiting the connection between internal fluxes and structure constants of the gaugings in the reduced supergravity theory, we can identify the types of fluxes arising in certain classes of freely-acting symmetric and asymmetric orbifolds. We give a general procedure for deriving the gauge algebra of the effective gauged supergravity using the exact CFT description at the orbifold point.  We find that the asymmetry is, in general, related to the presence of non-geometric $Q$- and $R$- fluxes. The action of T-duality is studied explicitly on various orbifold models and the resulting transformation of the fluxes is derived. Several explicit examples are provided, including compactifications with geometric fluxes,  $Q$-backgrounds (T-folds) and $R$-backgrounds. In particular, we present an asymmetric $\mathbb{Z}_4\times \mathbb{Z}_2$ orbifold in which all geometric  and non-geometric fluxes $\omega, H, Q, R$ are turned on simultaneously. We also derive the corresponding flux backgrounds, which are not in general T-dual to geometric ones, and may even simultaneously depend non-trivially on both the coordinates and their winding T-duals.
}
\end{abstract}
\end{center}

\smallskip}

\vfill
\begin{itemize}
\item[E-mails:] {\tt condeescu@roma2.infn.it}\,;\ \ {\tt florakis@mppmu.mpg.de}\,;\ {\tt kounnas@lpt.ens.fr}\,;\ \\ {\tt dieter.luest@lmu.de}.
\end{itemize}

\end{titlepage}
\end{spacing}

\tableofcontents


\section{Introduction}
\bigskip

String theory provides a framework in which the concepts of classical geometry are generalized in rather intriguing ways. These generalizations are very interesting not only due to  their rich mathematical structure, but also because they turn out to have important physical consequences. At short distances, namely at scales of the order of the string length, the very notion of Riemannian geometry is lost and is replaced by a new kind of stringy quantum geometry. One expects that this stringy geometry will eventually provide new insights into the physics of the Big Bang, into the nature of Black Holes and the resolution of space-time singularities. In fact, stringy geometry can be described by various different approaches. From the point of view of the string, classical geometry, dimensionality and even topology become effective notions, emerging only in the limit of low curvatures (low energy) (see e.g. \cite{Kiritsis:1994np}). The fully-fledged string theory is described in terms of an exact conformal field theory (CFT). In this context the background geometry in which the string propages is replaced by a two-dimensional field theory on the worldsheet, that is highly constrained by (super-) conformal invariance as well invariance under large (super-) reparametrizations  at all genera (modular invariance). Early examples of CFT constructions were given in terms  of bosonic covariant lattices \cite{Narain:1985jj}, fermionic constructions \cite{Antoniadis:1985az}, Gepner models \cite{Gepner:1987qi} as well as symmetric \cite{Dixon:1985jw} and asymmetric \cite{Narain:1986qm} orbifold CFT's. At length scales sufficiently larger than the string length, it is indeed sometimes possible to recover a geometric interpretation of the CFT constructions, e.g.
in terms of compact Calabi-Yau spaces or spaces with orbifold singularities. However, the description of stringy geometry in geometric terms is generically not possible.

Let us recall three important lessons that we have learned about stringy geometry so far. First, it became clear that duality symmetries play a decisive role in the study of stringy geometry. Contrary to the naive field theory intuition, seemingly different descriptions of string background geometries may turn out to be fully equivalent from the string perspective. This striking equivalence (duality) reflects the fact that the propagation of the string only depends on the underlying CFT, which may have different equivalent background realizations \cite{Kiritsis:1994np}. Indeed, consider the example of an $SU(2)_k$ WZW model which may be realized as an $S^3$ sphere at radius $R=\sqrt{k}$ with $k$ units of $H$-flux. At $k=1$ the system is equivalent to a $c=1$ system (free boson) on a circle $S^1$ at the self-dual radius.

The most prominent example of this equivalence among different classical geometries is the celebrated
T-duality symmetry.  It implies that there is no absolute (i.e. no invariant) notion of geometry or even topology in string theory. Indeed, it turns out that  a string can consistently propagate in non-geometric backgrounds, whose transition functions are not given
by standard diffeomorphisms but, rather, also involve non-trivial T-duality transformations. These spaces are often called T-folds \cite{Hull:2004in}, still being Riemannian manifolds locally; however, globally, these
spaces are characterized by non-trivial monodromies, where certain jumps in the  background fields, the metric $G_{MN}(X)$ and Kalb-Ramond field $B_{MN}(X)$, correspond to stringy symmetry operations.
This behavior can be formulated in terms  of non-geometric fluxes, namely in the case of T-folds, by the so-called
{\sl $Q$-fluxes.} These spaces themselves are often  T-dual to geometric spaces with geometric ``fluxes", namely {\sl $\omega$}- and/or {\sl $H$-fluxes.}
Even more dramatic are backgrounds with non-geometric {\sl $R$-fluxes,} which no longer admit a local description in terms of Riemannian spaces \cite{Shelton:2005cf,Dabholkar:2005ve}. Formally, $R$-flux spaces can be obtained from $Q$-flux backgrounds by applying T-duality transformations over non-isometry directions of the background.
As it became clear over the last few years, effective actions of T-folds and non-geometric spaces with $Q/R$-fluxes 
\cite{allp11,KLPT2007,Hassler:2013wsa}
can be  conveniently described
using the formalism of double geometry  and double field theory \cite{Hull:2009mi} (for reviews see \cite{Aldazabal:2013sca}). Non-geometric backgrounds can also arise as heterotic duals of F-theory constructions \cite{Becker:2009df}. In addition, there is a close relation between non-geometric string backgrounds and non-commutative and non-associative geometry \cite{Lust:2010iy,Condeescu:2012sp} (see also \cite{Chatzistavrakidis:2012qj} for a discussion in connection to matrix theory).

The second key observation, which will be important for this paper, is that closed strings can consistently propagate in asymmetric ``spaces" that look different for the left- and right-moving coordinates of the
 string. Asymmetric orbifolds constitute such classes of background CFT's. Moreover, a careful investigation
of non-geometric backgrounds with $Q$- or $R$-fluxes shows that these spaces also exhibit left-right asymmetry, in the sense that their monodromies act asymmetrically on the left- and right-moving string coordinates.
In fact, as was discussed in \cite{Dabholkar:2002sy,Hellerman:2002ax,Flournoy:2004vn,Hellerman:2006tx,Wecht:2007wu}, \cite{Condeescu:2012sp}, there is a close relation between asymmetric orbifolds and non-geometric string backgrounds. In particular, the consistent asymmetric orbifold CFT's constructed in \cite{Condeescu:2012sp} correspond to T-folds with non-geometric $Q$-flux at special points of moduli space. It is important to note that the relevant orbifold action is freely-acting; namely, a momentum shift along some particular coordinate $\mathbb{X}$, identified with the {\sl base} of the associated fibration, is accompanied by an (a)symmetric discrete rotation $\mathcal{M}$ of the remaining fiber coordinates $X^I$.

The third important point for our paper is the observation that geometric as well as non-geometric fluxes are closely related to the gauge algebra of the effective
supergravity theory, which is obtained after dimensional reduction on the associated geometric or non-geometric string backgrounds. In general, the corresponding effective
flux superpotentials \cite{der1,der2,der3}  become T-duality covariant functions only after including all possible $H,\omega,Q,R$-fluxes (see also \cite{Berman:2012uy} for reductions with U-duality twists).
Moreover, the superpotentials and the gauged supergravity algebra can be  derived from the double geometry formalism \cite{Blumenhagen:2013hva}, as well as from the intersection of
geometric and non-geometric branes \cite{Hassler:2013wsa}.
However, in this context, there still remains a partially unresolved puzzle.
Namely,  the gauged supergravity algebra  typically allows for a larger variety of simultaneously non-vanishing fluxes, compared to the number of fluxes that can actually be turned on when looking at specific (non)-geometric background spaces. For instance, the simultaneous appearance of $Q$- and $R$- fluxes, although  allowed
within the effective supergravity theory, could not be obtained so far by dimensional reduction on non-geometric spaces or  double field theory.

In this paper we elaborate on all three items mentioned above.
Concretely, the main points of discussion in our paper are as follows:

\begin{itemize}
\item The first part provides an explicit mapping between freely-acting orbifolds and fluxes. In Section \ref{GaugedSugraFluxes}, we explore the connection between (non-) geometric flux compactifications and their (a)symmetric orbifold limits, from the point of view of the effective gauged supergravity. In Section \ref{SecGaugingAlgebra}, we  derive in a  systematic way the algebra of gaugings including $H,\omega,Q,R$-fluxes from the vertex operator algebra of freely-acting asymmetric orbifolds. Recalling the standard construction of  geometric fluxes from left-right symmetric Scherk-Schwarz orbifolds, we show that asymmetric orbifold CFT's indeed provide  a very natural way of studying non-geometric spaces with $R$-flux at the full string level, setting them on equal footing with the spaces involving geometric or $Q$-fluxes. Indeed, whereas $Q$-fluxes originate from a momentum shift in the base together with an associated asymmetric $\mathcal{M}=\mathcal{M}_L\times \mathcal{M}_R$
group action in the fiber, we explicitly demonstrate that the T-dual $R$-fluxes are obtained from a winding shift in the base, accompanied by an asymmetric $ \mathcal{\tilde M}= \mathcal{\tilde M}_L\times
 \mathcal{\tilde M}_R$
group action in the fiber. Here the group elements $\mathcal{M}_L,\mathcal{\tilde M}_L$ act on the left-moving fiber coordinates $X^I_L$, whereas the $\mathcal{M}_R,\mathcal{\tilde M}_R$
 act on the right moving fiber coordinates $X^I_R$. Hence, a T-duality transformation along the base direction of the asymmetric orbifold CFT  maps $Q$- and $R$-fluxes into each other.
In addition, we will also see that the same T-duality  maps geometric $\omega$-fluxes into non-geometric $Q$-fluxes and vice versa. We further generalize this construction by considering combined momentum and winding shifts, which possess
a left-right asymmetric $\mathcal{M}\times \mathcal{\tilde M}$ group action in the fiber directions and we show that the resulting gauge algebra contains at the same time both non-geometric $Q$- and $R$-fluxes.

\item In the second part of the paper, we provide several explicit orbifold constructions that realize the general gauge algebras including geometric and non-geometric fluxes. In particular, in Section \ref{SecTdualityOrb}, we study the chain of T-dualities in the fiber and base directions, which connect the geometric flux to the $Q$- and $R$- flux frames. At the orbifold point the T-dualities we perform are exact at the string level, including in particular the T-duality in the base direction. Furthermore,  in Section \ref{ExampleSection} we present examples of  inherently non-geometric $Q$- and $R$- string backgrounds which cannot be T-dualized to geometric ones. In particular, we demonstrate the explicit realization of the above-mentioned combined momentum and winding shift and, hence, provide the proof of the conjecture \cite{Dabholkar:2005ve} for the existence of these backgrounds in string theory. Complementary to the orbifold CFT constructions, in Section \ref{BackgroundDesc} we also derive the target space background fields for various cases of interest. In this way, we also derive novel, more general  non-geometric T-fold backgrounds with $Q$- and even $R$-fluxes that are not T-dual to any geometric compactification, since the corresponding asymmetric orbifolds are also not T-dualizable to any symmetric orbifold construction. The method of providing  the map from the orbifold CFT's  to the T-folds with non-constant background fields $G,B$, relies in the faithful embedding of the discrete orbifold group $\mathcal{M}\times \mathcal{\tilde M}$ into the $O(d,d)$ duality group that acts on the background parameters $G,B$  of the fiber space. Using this information, one may derive the modular transformation rules of the fiber background fields, as one encircles the base coordinate ${\mathbb X}$.  As a result, in the case of $Q$-fluxes,
the background fields become periodic functions of the base coordinate ${\mathbb X}$, \emph{i.e.} $G({\mathbb X}),B({\mathbb X})$, whereas in the case of $R$-fluxes, they depend on the dual coordinates, $G(\tilde {\mathbb X}),B(\tilde {\mathbb X})$. Finally, in the more general case of combined momentum and winding shifts, the fiber background depends on both the base coordinate and its dual, $G({\mathbb X},\tilde {\mathbb X}),B({\mathbb X},\tilde {\mathbb X})$.
\end{itemize}


\section{Gauged supergravity and non-geometric fluxes}\label{GaugedSugraFluxes}

\subsection{Gaugings and fluxes}
In this section, we discuss the connection between gauged supergravity and compactifications with non-geometric fluxes. Let us first consider the Kaluza-Klein reduction of a ten dimensional theory consisting of a metric tensor $\mathcal{G}$, two-form field $\mathcal{B}$ and a scalar (dilaton) $\Phi$ described by the following action
\begin{align}
S=\int d^{10}x\sqrt{-\mathcal{G}}\, e^{-\Phi}\left[\mathcal{R}+(\nabla\Phi)^2-\frac{1}{12}\mathcal{H}^2\right]
\label{10d}\,,
\end{align}
where $\mathcal {R}$ and $\mathcal{H}$ are the ten-dimensional Ricci scalar and three-form field strength $\mathcal H = d \mathcal{B}$, respectively.
One can think of this as a subsector of Type II or Heterotic string theory; that is, the bosonic part of the $\mathcal{N}=1$ supergravity multiplet contained in each of them.
Compactification on a torus $T^D$ yields a reduced theory with $O(D,D)$ global symmetry and abelian $U(1)^{2D}$ gauge symmetry. The action of the reduced $(10-D)$-dimensional theory can be written in manifestly $O(D,D)$ invariant form
\begin{equation}
\begin{split}
S &=\int d^{10-D} x \sqrt{-g}\left[R+(\nabla \phi)^2-\frac{1}{12}H_{\mu\nu\rho}H^{\mu\nu\rho}\right.\\
&\left.\ +\frac{1}{8}L_{ab}\nabla_\mu K^{bc}L_{cd}\nabla^\mu K^{da}-\frac{1}{4}F_{\mu\nu}^aL_{ab}K^{bc}L_{cd}F^{d\mu\nu}\right]\, .
\end{split}\label{KK reduction}
\end{equation}
The Latin indices from the beginning of the alphabet $a,b,c,d$ running from $1$ to $2D$ are associated to the fundamental representation of the $O(D,D)$ global symmetry, whereas the Greek indices starting from $\mu,\nu, \rho = 1,...,10-D$ label the $10-D$ non-compact directions of space-time. The scalar fields in the reduced theory take values in the coset $\frac{O(D,D)}{O(D)\times O(D)}$ and are parametrized by a symmetric matrix $K^{ab}$, satisfying $\textrm{Tr}\, (L K) =0$, where $L$ is the standard $O(D,D)$ invariant metric
\begin{align}\label{oddmetric}
L=\left(
    \begin{array}{cc}
      0 & I \\
      I & 0 \\
    \end{array}
  \right)\,,
\end{align}
with $I$ being the D-dimensional unit matrix. The $2D$ gauge bosons $A^a_\mu=(V^M_\mu,B_{\mu M})$ arise from the reduction of the metric and of the B-field respectively, with the index $M=1,...,D$. If $\mathcal{M}\in O(D,D)$ is an arbitrary element of the group, then the corresponding transformation of the gauge bosons $A^a$ and of the scalars $K^{ab}$ is given by $A^a \rightarrow \mathcal{M}^a{}_b\, A^b$ and $K^{ab}\rightarrow \mathcal{M}^a{}_c\, \mathcal{M}^b{}_d \, K^{cd}$.

One can now gauge a $2D$ dimensional subgroup $\mathbb G \subset O(D,D)$. Denoting by $\mathcal{Z}_M$ the generators corresponding to the gauge bosons $V^M$ and by $\mathcal{X}^M$ the ones corresponding to $B_M$, one obtains a general gauge algebra  of the form \cite{Dabholkar:2005ve}:
\begin{align}
&[\mathcal{Z}_M,\mathcal{Z}_N]=\omega_{MN}^P\mathcal{Z}_P+H_{MNP}\mathcal{X}^P \,,\label{generalgauging1}\\
&[\mathcal{Z}_M,\mathcal{X}^N]= -\tilde\omega_{MP}^N\mathcal{X}^P+ Q^{NP}_M\mathcal{Z}_P \,,\label{generalgauging2}\\
&[\mathcal{X}^M,\mathcal{X}^N]= \tilde Q^{MN}_P\mathcal{X}^P+R^{MNP}\mathcal{Z}_P \,.\label{generalgauging3}
\end{align}
If the corresponding gauged supergravity follows from the (geometric) reduction of a higher dimensional supergravity theory, then the $2D$ vector fields follow from the reduction of the metric $G$ and $B$-field.
The structure constants $H, \omega, Q$ and $R$ have the interpretation of integrated geometric/non-geometric fluxes from the point of view of the compactification of a higher dimensional theory.
In geometric compactifications of ten dimensional supergravity one may only turn on the fluxes $H$ and $\omega$ and, thus, a general gauging containing also $Q$ and/or $R$ terms cannot be obtained by a geometric compactification of a higher dimensional supergravity theory. However, from the lower-dimensional point of view, such gaugings of supergravity do exist and can also be realized in the full string theory. For instance, a compactification with (elliptic) duality twists can be described at particular points of the moduli space by  a freely-acting asymmetric orbifold \cite{Condeescu:2012sp}, hence, providing a string realization of a $Q$-background. The presence of the $Q$-flux is associated to the asymmetry in the generalized fiber of the compactification. Furthermore, as we shall see, introducing an asymmetry also in the base, that is, considering an orbifold with asymmetric twists and shifts, yields a string realization of an $R$-background.


\subsection{Non-abelian gaugings from reductions of higher dimensional theories}
\subsubsection{Reduction on a twisted torus with flux (Scherk-Schwarz)}

Here we briefly consider the reduction of the action in eq. (\ref{10d}) on a twisted torus $T^D_{\omega}$ in the presence of three-form fluxes (for more details see \cite{Dabholkar:2005ve},\cite{Kaloper:1999yr}). These constructions are also known as Sherk-Schwarz compactifications \cite{Scherk:1978ta}. The internal manifold has a basis of one-forms $\eta^M$ such that the metric and B-field are given by
\begin{align}
&ds^2=G_{MN}\,\eta^M \eta^N\, , \qquad \qquad B=\frac{1}{2}B_{MN}\,\eta^M\wedge \eta^N+\varphi\, ,
\end{align}
such that $G_{MN}$ and $B_{MN}$ do not depend on the internal coordinates $X^M$ and, thus, give rise to $d^2$ scalar fields in the reduced theory.
Furthermore, the exterior derivatives of the one-forms $\eta^M$ and of the two-form $\varphi$ are
\begin{align}
d\eta^M = -\frac{1}{2}\omega_{NP}^M \,\eta^N\wedge\eta^P\, , \qquad \qquad d\varphi=-\frac{1}{3!}H_{MNP}\,\eta^M\wedge \eta^N \wedge \eta^P\, .
\end{align}
The compactification is determined by the constants $\omega _{NP}^M$ and $H_{MNP}$ which do not result in further moduli of the reduced theory. The twisted torus compactification is very similar to the Kaluza-Klein toroidal reduction considered in the previous section with the following replacement $dX^M \rightarrow \eta^M $. Notice that on the torus the twists $\omega$ are automatically zero since the exterior derivative is a nilpotent operator of order two $d^2=0$. However, in contrast to the case of toroidal Kaluza-Klein reduction, the gauge algebra for such a compactification becomes non-abelian. Explicitly, it is given by
\begin{align}
&[\mathcal{X}^M,\mathcal{X}^N]=0 \,,\label{ssgauge1}\\
&[\mathcal{Z}_M,\mathcal{X}^N]=- \omega_{MP}^N\mathcal{X}^P \,,\label{ssgauge2}\\
&[\mathcal{Z}_M,\mathcal{Z}_N]=\omega_{MN}^P\mathcal{Z}_P+H_{MNP}\mathcal{X}^P \,,\label{ssgauge3}
\end{align}
with the generators $\mathcal{Z}_M$ and $\mathcal{X}^M$ resulting from the reduction of the metric and the B-field, respectively. Notice that the symmetry generated by the gauge bosons coming from the B-field is always abelian in geometric compactifications.

\subsubsection{Reduction with T-duality twists (T-folds)}

We consider string theory compactified on a T-fold locally described by $T^{d}\times S^1$. The  coordinates on this space are then decomposed as follows $X^M = (X^I, \mathbb{X})$ with the coordinates $X^I$ associated to the ``fiber" and the coordinate $\mathbb{X}$ associated to the base space $S^1$. The theory reduced on $T^{d}$ has T-duality symmetry $O(d,d;\mathbb{Z})$. The further reduction on the circle further includes a twist by an element of the T-duality group. Explicitly, the fields are taken to depend on the circle coordinate $\mathbb{X}$ in the following way
\begin{align}
\psi(x^\mu,\mathbb{X})=\exp{\left(\frac{M \mathbb{X}}{2\pi R}\right)}\ \psi(x^\mu) \,,\label{twisted}
\end{align}
with $R$ being the radius of the circle $S^1$ and $\psi$ denoting an arbitrary field in the theory.
Hence, as $\mathbb{X}\rightarrow \mathbb{X}+2\pi R$, the fields acquire non-trivial monodromy, given by the monodromy matrix $\mathcal{M}$
\begin{align}
\mathcal{M}=\exp M\in O(d,d;\mathbb{Z})\,.
\end{align}
It is useful to introduce the following notation for the gauge generators
\begin{align}\label{twistredgenerators}
\mathcal{T}_a=(\mathcal{Z}_\mathbb{X},\mathcal{X}^\mathbb{X},\mathcal{T}_\alpha) \qquad,\qquad \mathcal{T}_\alpha=(\mathcal{Z}_I,\mathcal{X}^I)\,,
\end{align}
where the indices are $a=1,...,2D$, $\alpha=1,...,2d$ and $I=1,...,d$.
The gauge algebra for this reduction then takes the form
\begin{align}\label{twistedgauging}
[\mathcal{Z}_\mathbb{X},\mathcal{T}_{\alpha}]=M_{\alpha}{}^{\beta}\mathcal{T}_{\beta}\,,
\end{align}
with all other commutators vanishing. For elliptic monodromies the compactifications above admit (freely-acting, symmetric or asymmetric) orbifold  descriptions at particular points in the moduli space. Moreover, the orbifold point corresponds to a minimum of the scalar potential even in the asymmetric (elliptic) case. For parabolic monodromies, on the other hand, this is no longer true as, in general, these backgrounds do not admit orbifold fixed points, and their possible description in terms of  an exact CFT is highly non-trivial.

The form of the gauge algebra in eq. (\ref{twistedgauging}) will be derived in Section 3, by making use of the exact CFT description available at the orbifold point. The matrix $M$, determining the dependence on the internal circle coordinate $\mathbb{X}$ and encoding the fluxes present in the compactification, generates an order-$n$ rotation  $\mathcal{M}$ which, in turn,  precisely induces the action of the orbifold on the bosonic and fermionic worldsheet degrees of freedom (d.o.f.).

In order to compare with the general gauge algebra of eqs. (\ref{generalgauging1})-(\ref{generalgauging3}), it is convenient to parametrize $M$ in the basis $\mathcal{T}_\alpha=(\mathcal{Z}_I,\mathcal{X}^I)$ as
\begin{align}
M_\alpha{}^\beta=\left(
                   \begin{array}{cc}
                     W_I{}^J & U_{IJ} \\
                     V^{IJ} & -(W^t)^I{}_J \\
                   \end{array}
                 \right)\,,
\end{align}
where the $d\times d$ matrices satisfy $U_{IJ}=-U_{JI}$ and $V^{IJ}=-V^{JI}$ with $W_I{}^J$ unconstrained, as required in order for $M$ to be in the Lie algebra of $O(d,d)$.
By making use of eq. (\ref{twistedgauging}), the gauge algebra decomposes in the following way
\begin{align}
&[\mathcal{Z}_\mathbb{X},\mathcal{Z}_I]=W_I{}^J\mathcal{Z}_J+U_{IJ}\mathcal{X}^J\,,\\
&[\mathcal{Z}_\mathbb{X},\mathcal{X}^I]=-W_J{}^I\mathcal{X}^J+V^{IJ}\mathcal{Z}_J\,,\\
&[\mathcal{Z}_J,\mathcal{Z}_I]=0 \,,\qquad [\mathcal{X}^I,\mathcal{X}^J]=0\,,\\
&[\mathcal{X}^\mathbb{X}, \mathcal{Z}_M]=0 \,,\qquad [\mathcal{X}^\mathbb{X}, \mathcal{X}_M]=0\,,
\end{align}
where we have used the notation $\mathcal{Z}_M =(\mathcal{Z}_I, \mathcal{Z}_\mathbb{X}) $ and $\mathcal{X}^M=(\mathcal{X}^I, \mathcal{X}^\mathbb{X})$.
Notice that the gauge algebra above is of the same form as the one obtained in the case of Scherk-Schwarz compactifications  given in eqs. (\ref{ssgauge1})-(\ref{ssgauge3}) if and only if the matrix $V^{IJ}$ vanishes
\begin{align}
V^{IJ}=0\,.
\end{align}
In other words compactifications with non-zero $V^{IJ}$ are non-geometric. The fluxes may be easily identified by comparing with the general gauging in eqs. (\ref{generalgauging1})-(\ref{generalgauging3}). Indeed, the non-zero components of the fluxes are readily found to be
\begin{align}
\omega _{\mathbb{X} I}^J = \tilde \omega _{\mathbb{X} I}^J = W_I{}^J \, , \qquad H_{\mathbb{X}IJ} = U_{IJ}\, , \qquad Q_{\mathbb{X}}^{IJ}=V^{IJ}\, .
\end{align}
Even though these compactifications do not contain terms with non-trivial $\tilde Q^{MN}_P $ or $R^{MNP}$ fluxes, they can be ``truly'' non-geometric, in the sense that, for generic choices of $V,U$ and $W$, they are not T-dual to any geometric compactification. On the other hand, $\tilde Q$ or $R$ terms can arise by performing a T-duality in the base circle $S^1$.


\subsubsection{T-duality and non-geometric fluxes}
It is instructive to investigate the action of T-duality in connection with geometric and non-geometric fluxes
(see also the discussion in \cite{Hull:2009sg,Hull:2006qs}).
 Let us consider again the stringy compactification on a T-fold locally described by a space $T^d\times S^1$ with monodromy twist $\mathcal M \in O(d,d;\mathbb{Z})$. In particular, geometric fibrations can be described in this way, by making use of monodromies lying in the geometric subgroup $GL(d;\mathbb{Z}) \subset O(d,d;\mathbb{Z})$. The fluxes contained in such a compactification are encoded in the algebra generator of $\mathcal{M}$, parametrized by the matrix $M$. There are two kinds of T-dualities one may consider, depending on whether they act in the fiber or the base directions. The T-duality in the $T^d$ fiber is described by a matrix $\mathcal{O} \in O(d,d;\mathbb{Z})$ such that the fluxes contained in $M$ transform as
\begin{equation}
M' = \mathcal{O}^{} M \mathcal{O}^{-1}\,.
\end{equation}
Making use of the $O(d,d)$ invariant metric $L$ in eq. (\ref{oddmetric}), one can find the inverse of the T-duality matrix, $\mathcal{O}^{-1}=L \mathcal{O}^t L$. The metric $G(\mathbb{X})$ and Kalb-Ramond field $B(\mathbb{X})$, which depend on the base coordinate $\mathbb{X}$, transform according to the Buscher rules \cite{Buscher:1987sk}
\begin{align}
\mathcal{E}'(\mathbb{X}) = \bigr(A\mathcal{E}(\mathbb{X})+B\bigr)\bigr(C\mathcal{E}(\mathbb{X})+D\bigr)^{-1}\,,
\end{align}
where we defined $\mathcal{E}(\mathbb{X})\equiv G(\mathbb{X})+B(\mathbb{X})$ and the T-duality matrix $\mathcal{O}$ and its inverse $\mathcal {O}^{-1}$ are parametrized by
\begin{equation}
\mathcal{O}=\left(
              \begin{array}{cc}
                A & B \\
                C & D \\
              \end{array}
            \right)\,,
\qquad
\mathcal{O}^{-1}=\left(
              \begin{array}{cc}
                D^t & B^t \\
                C^t & A^t \\
              \end{array}
            \right)\,.
\end{equation}
The $d$-dimensional matrices $A,B,C,D$ are subject to the constraints
\begin{equation}
A^tC+C^tA=0\, ,\quad B^tD+D^tB=0\, ,\quad A^tD+C^tB=I\, .
\end{equation}
For simplicity, let us consider the case of a geometrically fibered space with monodromy generated by the following flux matrix $M$
\begin{equation}\label{blockdiagonal}
M=\left(
    \begin{array}{cc}
      W & 0 \\
      0 & -W^t \\
    \end{array}
  \right)\,,
\end{equation}
parametrized as before in the basis $\mathcal{T}_\alpha=(\mathcal{Z}_I,\mathcal{X}^I)$ with off-diagonal elements $U=V=0$. The only fluxes present in this compactification are $\omega$ and $\tilde\omega$. After performing the T-duality $\mathcal{O}$, one arrives at the  transformed flux matrix
\begin{equation}
M'= \left(
    \begin{array}{cc}
      AWD^t-BW^tC^t & AWB^t-BW^tA^t \\
      CWD^t-DW^tC^t & CWB^t-DW^tA^t \\
    \end{array}
  \right)\,.
\end{equation}
In the new duality frame, we now have both geometric and non-geometric fluxes present, $H,\omega, \tilde \omega$ and $Q$. However this is not a ``true'' Q-background as it is T-dual to a geometric one. In view of the above, we will call ``true" Q-backgrounds, those ones for which the matrix $M$ generating the monodromy $\mathcal{M}$ of the compactification does not belong to the conjugacy class of $M_{geom}$, defined as
\begin{equation}
M_{geom} = \left(
    \begin{array}{cc}
      W & U \\
      0 & -W^t \\
    \end{array}
  \right)\ .
\end{equation}
Notice that the rules for performing T-duality used in the arguments above can be justified at the level of the gauged $\sigma$-model \cite{Buscher:1987sk}. This is no longer the case when one tries to perform a T-duality in the base direction. The reason is that, in this case, there is no longer an isometry that one can gauge, due to the explicit dependence of the background fields on $\mathbb{X}$. Hence, for a fully-fledged flux compactification, the T-duality in the base cannot be performed in the usual way. However, at the orbifold point, because the dependence on $\mathbb{X}$ enters only through boundary conditions, the T-duality can still be carried out exactly at the CFT level and a subsequent deformation away from the orbifold point could be used to define the new background consistently. In this way, the gauge algebra in eq. (\ref{twistedgauging}) becomes
\begin{align}
[\mathcal{X}^{\mathbb{X}},T_{\alpha}]=M_{\alpha}{}^{\beta} T_{\beta}\, ,
\end{align}
effectively interchanging the generators $\mathcal{Z}_{\mathbb{X}} \leftrightarrow \mathcal{X}^{\mathbb{X}}$. Notice that, starting from a compactification described by $M_{geom}$ and performing a T-duality in $S^1$, the resulting non-vanishing fluxes are $\tilde \omega, Q, \tilde Q $. Therefore, in order to obtain an $R$-flux background, one needs to start from  a non-geometric fiber with $Q$-flux (\emph{i.e.} work in the non-geometric $Q$-frame). Namely, starting with a generic matrix $M$ with all $W,V,U \neq 0$, the T-duality in the base direction leads to a compactification with $\tilde \omega, Q, \tilde Q , R$ fluxes. Specifically, the fluxes are mapped as follows
\begin{equation}\label{mappingfluxes}
		H \rightarrow \tilde\omega \,,\quad \omega\rightarrow Q \,,\quad \tilde\omega\rightarrow\tilde{Q}\,,\quad Q\rightarrow R\,.
\end{equation}
In the next section we investigate ``true" $R$-backgrounds with asymmetric orbifold descriptions.


\subsubsection{R-backgrounds from orbifolds with asymmetric twists and shifts}

Freely-acting asymmetric orbifolds provide a powerful tool for studying the properties of compactifications with non-geometric fluxes. With this in mind, one may construct ``true" R-backgrounds (i.e. where the R-flux cannot be eliminated by any T-duality) as orbifolds with asymmetric twists and shifts. Finding well-defined asymmetric actions is, however, a non-trivial task, due to the difficulty in satisfying the  constraints of (multi-loop) modular invariance \cite{Narain:1986qm}, \cite{Aoki:2004sm} (see also \cite{Bianchi:1999uq} for examples of asymmetric constructions). As we have seen in the previous section, the presence of Q-flux was related to the asymmetry in the $T^{d}$ ``fiber". A further generalization of this, is to introduce an asymmetry also in the base space $S^1$, in the form of asymmetric shifts in $\mathbb{X}_L,\mathbb{X}_R$. Concretely, we consider an orbifold of the form $\mathbb{Z}_n\times \mathbb{Z}_m$ with the matrix $M$ generating the rotation of order $n$ and the matrix $\tilde M$ similarly generating the order-$m$ rotation. Furthermore, the matrices $M$ and $\tilde M$ are taken to commute with one another. The action of $\mathbb{Z}_n$ is accompanied by a shift in the coordinate $\mathbb{X}$, whereas the action of $\mathbb{Z}_m$ is accompanied by a shift in the dual coordinate $ \mathbb{\tilde X}$, thus ensuring the asymmetry of the base. Performing a T-duality in the base $S^1$ then effectively interchanges the two orbifold rotations
\begin{equation}
\mathbb{Z}_n\ \stackrel{T}{\longleftrightarrow}\ \mathbb{Z}_m\,.
\end{equation}
The gauge algebra in this case will have the form \cite{Dabholkar:2005ve}
\begin{align}
&[\mathcal{Z}_\mathbb{X},\mathcal{T}_{\alpha}]=M_{\alpha}{}^{\beta}\mathcal{T}_{\beta}\,, \label{Rgauging1}\\
&[\mathcal{X}^\mathbb{X},\mathcal{T}_\alpha]=\tilde M_{\alpha}{}^{\beta}\mathcal{T}_{\beta}\,, \label{Rgauging2}
\end{align}
which suggests the following non-local dependence of the fields on the internal non-geometric circle~$S^1$
\begin{align}\label{momwindreduction}
\psi(x^\mu, \mathbb{X}, \mathbb{\tilde X})=\exp{\left(\frac{M \mathbb{X}}{2\pi R}\right)}\exp{\left(\frac{\tilde M \mathbb{\tilde X}}{2\pi \tilde R}\right)}\ \psi(x^\mu)\,.
\end{align}
The dependence of the  fields on both $\mathbb{X}$ and $\tilde{\mathbb{X}}$  is a generic feature of ``true" $R$-backgrounds (see also the discussion in Section \ref{Z4Z2backgr}). The above equation was proposed in \cite{Dabholkar:2005ve} and we shall argue its validity from the CFT derivation of eqs. (\ref{Rgauging1}), (\ref{Rgauging2}) in Section \ref{SecZnZmAlg}. It is again instructive to decompose the algebra above in the basis $\mathcal{T}_\alpha=(\mathcal{Z}_I,\mathcal{X}^I)$, according to
\begin{align}
M_\alpha{}^\beta=\left(
                   \begin{array}{cc}
                     W_I{}^J & U_{IJ} \\
                     V^{IJ} & -(W^t)^I{}_J \\
                   \end{array}
                 \right)\,,
\qquad
\tilde M_\alpha{}^\beta=\left(
                   \begin{array}{cc}
                     \tilde W_I{}^J & \tilde U_{IJ} \\
                     \tilde V^{IJ} & -(\tilde W^t)^I{}_J \\
                   \end{array}
                 \right)\,.
\end{align}
Making use of the above parametrization, the non-zero commutators are found to be
\begin{align}
&[\mathcal{Z}_\mathbb{X},\mathcal{Z}_I]=W_I{}^J\mathcal{Z}_J+U_{IJ}\mathcal{X}^J \,,\\
&[\mathcal{Z}_\mathbb{X},\mathcal{X}^I]=-W_J{}^I\mathcal{X}^J+V^{IJ}\mathcal{Z}_J \,,\\
&[\mathcal{X}^\mathbb{X},\mathcal{Z}_I]=\tilde W_I{}^J\mathcal{Z}_J+\tilde U_{IJ}\mathcal{X}^J \,,\\
&[\mathcal{X}^\mathbb{X},\mathcal{X}^I]=-\tilde W_J{}^I\mathcal{X}^J+\tilde V^{IJ}\mathcal{Z}_J \,.
\end{align}
One may then compare with in eqs. (\ref{generalgauging1})-(\ref{generalgauging3}) in order to obtain the explicit identification of the fluxes
\begin{align}
&\omega _{\mathbb{X} I}^J = \tilde \omega _{\mathbb{X} I}^J = W_I{}^J \, , \qquad H_{\mathbb{X}IJ} = U_{IJ}\, , \qquad Q_{\mathbb{X}}^{IJ}=V^{IJ}\, , \\
&\tilde \omega _{ IJ }^\mathbb{X} =- \tilde U_{IJ} \, ,  \qquad Q_{I}^{\mathbb{X}J}= - \tilde Q_{I}^{\mathbb{X}J}=- \tilde{W}_{I}{}^J\, , \qquad R^{\mathbb{X}IJ} = \tilde V^{IJ}  \, ,
\end{align}
It is straightforward to see that in these models the $R$-flux terms in the gauge algebra cannot be T-dualized away. These compactifications are quite general, since they contain all the fluxes simultaneously. We shall provide an explicit example for this type of background, based on a $\mathbb{Z}_4 \times \mathbb{Z}_2$ orbifold in Section \ref{Z2Z4orbCFT}.



\section{Freely-acting orbifolds and gauged supergravity}\label{SecGaugingAlgebra}

In order to obtain a systematic identification of the fluxes involved in freely-acting (a)symmetric orbifold models, it is important to make contact with the effective supergravity description. The integrated fluxes can be described \cite{der1,der2,der3} as gaugings of the supergravity theory  and, in this section, we derive the corresponding gauge algebra for a generic class of freely-acting (a)symmetric orbifolds\footnote{Examples of freely-acting asymmetric orbifolds and their relation to ``gravito-magnetic'' fluxes have also been discussed \emph{e.g.} in \cite{Angelantonj:2008fz} in the context of string thermodynamics and string cosmology. In the latter context, a T-fold description of a parafermionic CFT was presented in \cite{Kounnas:2007pg}.}. In particular, we will not limit ourselves to orbifolds which are connected to symmetric ones by a chain of T-dualities, but rather consider quite generic cases which may lie in different conjugacy classes of the T-duality group, than the symmetric ones. These are the ``truly'' non-geometric backgrounds which cannot be cast into a geometric frame by any T-duality transformation and for which explicit examples will be given in Section \ref{ExampleSection}.


\subsection{Generic shift in the base}\label{SecGeneralShifts}

Let us start by discussing the structure of the generic freely-acting orbifolds we will consider, from the point of view of the base. For simplicity, we will focus on freely-acting $(T^d\times S^1)/ \mathbb{Z}_N$ orbifolds, whose action takes the form:
\begin{equation}
	\mathcal{G} = e^{2\pi (F_L+ F_R)}\, \delta_{a,b} \,, \label{orbAction}
\end{equation}
where $\delta_{a,b}$ is an order-$N$  shift along the base $S^1$ and $F_{L,R}$ are the generators of the orbifold rotation on the fiber coordinates $X^I$. The asymmetry in the fiber is then characterized by $F_L,F_R$, with $F_L\neq F_R$ corresponding to an asymmetric action. Of course, we assume that the choice of the orbifold action in eq. \eqref{orbAction} corresponds to a well-defined string vacuum, consistent with the constraints of modular invariance and unitarity.

We now focus on the base direction $\mathbb{X}$ and its action under the generic shift $\delta_{a,b}$. Its partition function is given by a shifted lattice sum $\Gamma_{(1,1)}[{H\atop G}]$ of order $N$ which, in the Hamiltonian representation, reads:
\begin{equation}\label{latticeHamiltonian}
	\sum\limits_{m,n\in\mathbb{Z}} e^{2\pi i G\frac{a m+ b n}{N}} \exp\left[\frac{i\pi\tau}{2}\left(\frac{m+b \frac{H}{N}}{R}+\left(n+ a \tfrac{H}{N}\right)R\right)^2 - \frac{i\pi\bar\tau}{2}\left(\frac{m+b \frac{H}{N} }{R}-\left(n+ a \tfrac{H}{N}\right)R\right)^2\right] \,.
\end{equation}
The shift\footnote{This can be seen from the vertex operator $\mathcal{V}=\exp(iP_L \mathbb{X}_L+iP_R \mathbb{X}_R)$ contribution to the lattice sum.} along $S^1$ is parametrized by the vector $\lambda = (\frac{a}{N},\frac{b}{N})$, with $a,b$ defined modulo $N$:
\begin{equation}
	\delta_{a,b} \ :\  \left\{  \begin{split}
		& \mathbb{X}_L \rightarrow \mathbb{X}_L + a \frac{\pi R}{N} + b \frac{\pi}{NR} \\
		& \mathbb{X}_R \rightarrow \mathbb{X}_R + a \frac{\pi R}{N} - b \frac{\pi}{NR}\\
	\end{split}\right.  \ .
\end{equation}
or, in terms of the circle coordinate and its dual:
\begin{equation}
	\delta_{a,b} \ :\  \left\{  \begin{split}
		& \mathbb{X} \rightarrow \mathbb{X} + a \frac{2\pi R}{N}  \\
		& \tilde{\mathbb{X}} \rightarrow \tilde{\mathbb{X}} + b \frac{2\pi}{NR}\\
	\end{split}\right.  \ .
\end{equation}
The case $(a,b)=(1,0)$ clearly shifts the coordinate $\mathbb{X}$, while leaving its dual $\tilde{\mathbb{X}}$ invariant and corresponds to what is called a momentum shift\footnote{Notice that the terminology momentum/winding shift is slightly counter-intuitive from the point of view of the Hamiltonian lattice eq.\eqref{latticeHamiltonian}. Indeed, inspecting the physical mass, a momentum shift $a=1,b=0$ is characterized by an $H$-shift in the winding number $n$, whereas a winding shift $a=0,b=1$ is accompanied by an $H$-shift in the momentum quantum number $m$. This terminology has its origin in the corresponding Scherk-Schwarz picture, where the freely-acting orbifold is represented as a momentum or winding boost in the charge lattices, as discussed in the following sections.}, recognized by the phase $\exp(2\pi i m /N)$ in the Hamiltonian representation of the lattice in eq.\,\eqref{latticeHamiltonian}. This is the case for the geometric (Scherk-Schwarz) compactification.
Under $T$-duality in the base $S^1$, one obtains $(a,b)=(0,1)$ which shifts the dual coordinate $\tilde{\mathbb{X}}$, while preserving $\mathbb{X}$ unshifted. This is the case of the winding shift, recognized by the phase $\exp(2\pi i n/N)$ in the Hamiltonian lattice, and we will show in the following sections that, whenever the fiber also carries asymmetric monodromy, it corresponds to the stringy origin of the  $R$-flux.

Finally, there is the more interesting case of combined momentum \emph{and} winding shift, where both the coordinate and its dual are shifted. It  corresponds to ``inherently'' asymmetric constructions for which the asymmetry on the base cannot be geometrized by any T-duality action.  They will be shown in the following sections to correspond to ``truly'' non-geometric string backgrounds involving both $Q$- and $R$- fluxes.


\subsection{Momentum shift, the algebra of gaugings and $Q$-flux}

We now address the problem of determining the algebra of gaugings for a consistent freely-acting  $\mathbb{Z}_N$-orbifold with pure momentum shift:
\begin{equation}
	\mathcal{G} = e^{2\pi (F_L+ F_R)}\, \delta_{1,0} \,.
\end{equation}
The starting point relies on the key observation \cite{Kounnas:1988ye} that the above orbifold action can be realized as an  $\frac{O(d,d)}{O(d)\times O(d)}$ boost in the fermionic and bosonic charge lattices by a $\mathbb{Z}_N$-quantized boost parameter. We focus on the fermionic charge lattice, together with the $S^1$ lattice of the base:
\begin{equation}
	\left\{\begin{split}
		& Q_L^i \rightarrow Q_L^i - \xi_L^i (P_L^0-P_R^0) \\
		& Q_R^i \rightarrow Q_R^i -\xi_R^i (P_L^0-P_R^0) \\
		& P_L^0 \rightarrow P_L^0 + \xi_L\cdot Q_L-\xi_R\cdot Q_R - \tfrac{1}{2}(\xi_L^2-\xi_R^2) (P_L^0-P_R^0) \\
		& P_R^0 \rightarrow P_R^0 + \xi_L\cdot Q_L - \xi_R \cdot Q_R - \tfrac{1}{2} (\xi_L^2-\xi_R^2) (P_L^0-P_R^0)\\
	\end{split}\right. \ , \label{boost}
\end{equation}
where $\xi_{L,R}^I$ are the quantized left- and right- moving boost parameters and the index $i$ labels the (left- or right- moving) chiral bosons $H^i$ arising after the bosonization of the (complexified) worldsheet fermions in the internal directions. The generic vertex operator for the ground states of the theory contains the internal part:
\begin{equation}
	\mathcal{V}(z,\bar{z}) =  e^{i Q_L\cdot H_L(z) + i Q_R\cdot H_R(\bar{z})}\ e^{iP_L^0 \mathbb{X}_L(z)+iP_R^0\mathbb{X}_R(\bar z)} \,.
\end{equation}
On this generic vertex operator the boost \eqref{boost} acts as:
\begin{equation}
	\mathcal{V}(z,\bar{z}) \rightarrow \exp\left[i \mathcal{Q} (\mathbb{X}_L+ \mathbb{X}_R)\right] \ \hat{\mathcal{V}}(z,\bar{z}) \,,
\end{equation}
where:
\begin{equation}
	\mathcal{Q} = \xi_L\cdot Q_L - \xi_R \cdot Q_R -\tfrac{1}{2}(\xi_L^2-\xi_R^2)(P_L^0-P_R^0) \,,
\end{equation}
and:
\begin{equation}
	\hat{\mathcal{V}}(z,\bar{z}) = \exp\left[-i (P_L^0-P_R^0)\left(\xi_L\cdot H_L+\xi_R\cdot H_R\right)\right] \mathcal{V}(z,\bar{z}) \,.
\end{equation}
The relation to the Scherk-Schwarz reduction in field theory can be seen immediately by considering the supergravity limit of the above equations for the associated gauge bosons. For these states, $P_L^0+P_R^0\sim m/R=0$, $P_L^0-P_R^0 \sim nR =0$ and the boost has the simple effect:
\begin{equation}
	\mathcal{V}(z,\bar{z}) \rightarrow \exp\left[i \mathcal{Q}\, \mathbb{X}(z,\bar{z})\right] \ \mathcal{V}(z,\bar{z}) \,.
\end{equation}
For the ground states present in the effective supergravity, one may consider $Q_L^i,Q_R^i$ to be identified with the charges  of the left- and right- moving worldsheet fermions $\psi^I_L, \psi^I_R$, respectively. Notice that a state $\mathcal{V}_\alpha$ with definite $Q_L, Q_R$ charges will be boosted accordingly, with boosting parameters $\xi_L, \xi_R$. Upon a state which is not necessarily an eigenmode, the boost acts as:
\begin{equation}
	\mathcal{V}_\alpha(z,\bar{z}) \rightarrow \left[e^{i (\xi_L \cdot Q_L - \xi_R \cdot Q_R) \mathbb{X}}\right]_{\alpha\beta} \ \mathcal{V}_\beta(z,\bar{z}) \,.\label{SchSchMom}
\end{equation}
Hence, there is a one-to-one map between the transformation matrices $[e^{i (\xi_L \cdot Q_L - \xi_R \cdot Q_R)}]_{\alpha\beta}$ and the $2d$ gauge bosons $\{e^{iQ_L\cdot H_L+iQ_R\cdot H_R}\}=\{\psi^\mu\tilde\psi^I\oplus\psi^I\tilde\psi^\mu\}$ of the toroidal $O(d,d)$ reduction of supergravity, determined by their helicity charges $Q_L,Q_R$. As a result, for a given (quantized) choice of the boosting parameters $\xi_L,\xi_R$, specified by the particular orbifold in question, eq. \eqref{SchSchMom} precisely corresponds to the field-theoretic Scherk-Schwarz reduction in eq. \eqref{twisted}.

At this stage, it will be instructive to provide a stringy derivation of eq. \eqref{twistedgauging}, that admits a straightforward generalization to the more general cases that we will encounter in the following sections. To this end, we define the $U(1)$ charges associated to the $S^1$ base:
\begin{equation}
	\left\{\begin{split}
	&Q^0=\tfrac{1}{2}(Q^0_L+Q^0_R)  \\
	&\tilde{Q}^0 = \tfrac{1}{2}(Q^0_L - Q^0_R) \\
	\end{split}\right.
\quad,\quad
	\left\{\begin{split}
	&Q^0_L = \oint \frac{dz}{2\pi}~\partial \mathbb{X}  \\
	&Q^0_R = -\oint \frac{d\bar{z}}{2\pi}~\bar\partial \mathbb{X}
	\end{split}\right. \, .
\end{equation}
These are precisely identified with the generators $\mathcal{Z}_\mathbb{X}\equiv Q^0, \mathcal{X}^\mathbb{X}\equiv \tilde{Q}^0$ of eq. \eqref{twistredgenerators}. Their action on the remaining gauge bosons is then readily obtained using the operator product expansion (OPE) of the above vertex operators:
\begin{equation}
	\begin{split}
	&[\mathcal{Z}_{\mathbb{X}}, \mathcal{T}_\alpha] = \tfrac{1}{2}\left[\oint \frac{dz}{2\pi}~ \partial \mathbb{X} - \oint \frac{d\bar{z}}{2\pi}~\bar\partial \mathbb{X} \right]\mathcal{T}_\alpha= (\xi_L\cdot Q_L-\xi_R\cdot Q_R)_{\alpha\beta} \ \mathcal{T}_\beta  \\
	&[\mathcal{X}^{\mathbb{X}}, \mathcal{T}_\alpha] = \tfrac{1}{2}\left[\oint \frac{dz}{2\pi}~ \partial \mathbb{X} + \oint \frac{d\bar{z}}{2\pi}~\bar\partial \mathbb{X} \right]\mathcal{T}_\alpha= 0
	\end{split}\,. \label{algmomentum}
\end{equation}
Notice that a tensor sum is assumed in the r.h.s. of the above equation\footnote{Similarly, one may also derive the commutator $[T_\alpha,T_\beta]$ using the OPEs of the massive gauge bosons after the boost (necessarily with opposite charge $Q$). However, it turns out that they produce higher oscillator states carrying masses of the order of the string scale and, hence, we can effectively truncate the algebra by assuming that $[T_\alpha,T_\beta]=0$ in the supergravity limit.}, since $Q_L$ and $Q_R$ act on the diagonal $O(d)_{L}\times O(d)_R\subset O(d,d)$.

Furthermore, it is clear from eq. \eqref{SchSchMom} that a shift in $\mathbb{X}$ induces precisely the orbifold action on the remaining worldsheet d.o.f. . Hence, we can  parametrize them in terms of the flux matrices $F_L,F_R$, generating the left- and right- moving boosts:
\begin{equation}\label{ident1}
	F_L =  \xi_L\cdot Q_L \quad, \quad F_R = - \xi_R\cdot Q_R \,.
\end{equation}
 In order to obtain the precise match between the left- and right- moving flux matrices $F_L, F_R$ on the one hand, and the flux matrix $M$ of the twisted reduction in eq. \eqref{twisted} on the other, we simply need to express $F_{L,R}$ in terms of the $(X^I, \tilde{X}^I)$-basis of the fiber coordinates and their duals
\begin{equation}\label{ident2}
	M = U^{-1}\, \mathbb{F} \ U = \frac{1}{2} \left(\begin{array}{c c}
									F_L+F_R & F_L-F_R \\
									F_L-F_R & F_L+F_R\\
									\end{array}\right)	\,,
\end{equation}
where $\mathbb{F} = F_L\oplus F_R$ is the tensor sum of the left- and right- moving flux matrices and:
\begin{equation}
	U = \tfrac{1}{2}\left( \begin{array}{c r}
						I & I \\
						I & -I\\
						\end{array}\right)~,
\end{equation}
is the $2d\times 2d$ matrix taking us from the basis $(X^I,\tilde{X}^I)$ to the basis $(X_L^I,X_R^I)$. Explicitly, the gauge algebra takes the form:
\begin{equation}
\begin{split}
&[\mathcal{Z}_\mathbb{X},\mathcal{Z}_I]=\tfrac{1}{2}(F_L+F_R)_I{}^J \mathcal{Z}_J+\tfrac{1}{2}(F_L-F_R)_{IJ}\mathcal{X}^J\\
&[\mathcal{Z}_\mathbb{X},\mathcal{X}^I]=-\tfrac{1}{2}(F_L+F_R)_J{}^I \mathcal{X}^J+\tfrac{1}{2}(F_L-F_R)^{IJ}\mathcal{Z}_J\\
\end{split}\,. \label{AlgMomentumShift}
\end{equation}
Comparing with eqs. \eqref{generalgauging1}-\eqref{generalgauging3}, one may read off  the presence of $Q$-flux, provided the action of the orbifold on the fiber is asymmetric, $F_L\neq F_R$.


\subsection{Winding shift  and R-flux}

We now move on to the T-dual case with respect to the base, corresponding to a freely-acting orbifold with winding shift:
\begin{equation}
	\mathcal{G} = e^{2\pi (F_L+ F_R)}\, \delta_{0,1} \,.
\end{equation}
In this case, it is straightforward to see that the boost takes the T-dual form:
\begin{equation}
	\left\{\begin{split}
		& Q_L^i \rightarrow Q_L^i - \xi_L^i (P_L^0+P_R^0) \\
		& Q_R^i \rightarrow Q_R^i -\xi_R^i (P_L^0+P_R^0) \\
		& P_L^0 \rightarrow P_L^0 + \xi_L\cdot Q_L-\xi_R\cdot Q_R - \tfrac{1}{2}(\xi_L^2-\xi_R^2) (P_L^0+P_R^0) \\
		& P_R^0 \rightarrow P_R^0 - (\xi_L\cdot Q_L - \xi_R \cdot Q_R) + \tfrac{1}{2} (\xi_L^2-\xi_R^2) (P_L^0+P_R^0)\\
	\end{split}\right. \ . 
\end{equation}
As in the previous section, the action of the boost on the vertex operators is given by:
\begin{equation}
	\mathcal{V}(z,\bar{z}) \rightarrow \exp\left[i \mathcal{Q} (\mathbb{X}_L- \mathbb{X}_R)\right] \ \hat{\mathcal{V}}_\alpha(z,\bar{z}) \,,
\end{equation}
where the operator $\mathcal{Q}$, associated to the fermionic helicity charges, now becomes:
\begin{equation}
	\mathcal{Q} = \xi_L\cdot Q_L - \xi_R \cdot Q_R -\tfrac{1}{2}(\xi_L^2-\xi_R^2)(P_L^0+P_R^0) \,,
\end{equation}
and:
\begin{equation}
	\hat{\mathcal{V}}(z,\bar{z}) = \exp\left[-i (P_L^0+P_R^0)\left(\xi_L\cdot H_L+\xi_R\cdot H_R\right)\right] \mathcal{V}(z,\bar{z}) \,.
\end{equation}
Again, for the gauge bosons, $P_L^0+P_R^0\sim m/R=0$, $P_L^0-P_R^0\sim nR=0$, so that the Scherk-Schwarz boost becomes:
\begin{equation}
	\mathcal{V}_\alpha(z,\bar{z}) \rightarrow \left[e^{i (\xi_L \cdot Q_L - \xi_R \cdot Q_R) \tilde{\mathbb{X}}}\right]_{\alpha\beta} \ \mathcal{V}_\beta(z,\bar{z}) \,,
\end{equation}
 now involving the T-dual coordinate $\tilde{\mathbb{X}}=\mathbb{X}_L-\mathbb{X}_R$ of the base. Working in a similar way as in the case of the momentum shift, one may derive the algebra of gaugings of the effective supergravity using the OPEs:
\begin{equation}
	\begin{split}
	&[\mathcal{Z}_\mathbb{X}, \mathcal{T}_\alpha] =0 \\
	&[\mathcal{X}^\mathbb{X}, \mathcal{T}_\alpha]  = (\xi_L\cdot Q_L-\xi_R\cdot Q_R)_{\alpha\beta} \ \mathcal{T}_\beta
	\end{split}\,. \label{algwinding}
\end{equation}
Explicitly, in terms of the flux matrices $F_L,F_R$ parametrizing the action of the orbifold on the fiber, the gauge algebra takes the form:
\begin{equation}
\begin{split}
&[\mathcal{X}^\mathbb{X},\mathcal{Z}_I]=\tfrac{1}{2}(F_L+F_R)_I{}^J \mathcal{Z}_J+\tfrac{1}{2}(F_L-F_R)_{IJ}\mathcal{X}^J\\
&[\mathcal{X}^\mathbb{X},\mathcal{X}^I]=-\tfrac{1}{2}(F_L+F_R)_J{}^I \mathcal{X}^J+\tfrac{1}{2}(F_L-F_R)^{IJ}\mathcal{Z}_J\\
\end{split}\,.\label{AlgWindingShift}
\end{equation}
Comparing with eqs. \eqref{generalgauging1}-\eqref{generalgauging3}, one may identify in this gauging the presence of non-zero $R$-flux, provided the action of the orbifold on the fiber is also asymmetric, $F_L\neq F_R$.


\subsection{Simultaneous momentum and winding shift}\label{SecAlgMomentumWindingShift}

Let us now consider the more general case of a freely-acting asymmetric orbifold where the shift in the base is also asymmetric, corresponding to a combined shift in both momenta and windings:
\begin{equation}
	\mathcal{G} = e^{2\pi (F_L+ F_R)}\, \delta_{1,1} \,.
\end{equation}
In this case, due to the simultaneous momentum and winding shift, the asymmetry of the base induces an irreducible non-locality into the background and has drastic effects on the algebra of gaugings. Here, it will be more convenient to work directly in the orbifold representation.

We focus on the states corresponding to the gauge bosons before the gauging. These lie in the untwisted sector $h=0$ and come with definite fermion charges $Q_L,Q_R$. In the partition function, the boundary conditions of the fermions are twisted as:
\begin{equation}
	\prod_{i-\textrm{left}}\vartheta\biggr[{\alpha_i-2\xi_L^i H \atop \beta_i-2\xi_L^i G}\biggr] \ \prod_{j-\textrm{right}}\bar\vartheta\biggr[{\alpha_j-2\xi_R^j H \atop \beta_j-2\xi_R^j G}\biggr]\,,
\end{equation}
where $\alpha_i,\beta_i$ correspond to the spin-structures. Summation over $G\in\mathbb{Z}_N$ then projects onto the states invariant under the orbifold and yields the constraint:
\begin{equation}
	m+n=  N\xi\cdot Q \ (\textrm{mod}\ N)\,,
\end{equation}
where $\xi\cdot Q\equiv \xi_L\cdot Q_L-\xi_R\cdot Q_R$ and $m,n\in\mathbb{Z}$ are the momentum and winding quantum numbers, respectively. Furthermore, level matching of the $S^1$-lattice imposes the constraint $(P_L^0)^2-(P_R^0)^2\sim mn=0$, since the gauge bosons in the original (ungauged) theory already come level-matched with fermionic oscillator weight $(\Delta,\bar\Delta)=(\frac{1}{2},\frac{1}{2})$. The solution of the above constraints takes the form:
\begin{equation}
	A^\alpha_\mu\leftrightarrow\{ m=N\xi\cdot Q \ , \ n=0 \} \qquad,\qquad \tilde{A}^\alpha_\mu\leftrightarrow\{ m=0 \ , \ n= N\xi\cdot Q \} \,,
\end{equation}
where $A^\alpha_\mu$ and $\tilde{A}^\alpha_\mu$ are the gauge bosons carrying non-trivial momentum and winding charge, respectively. The above non-trivial charges completely fix the algebra of gaugings. The gauge bosons in the gauged theory acquire a mass, since their internal part now involves the $S^1$ contribution:
\begin{equation}
	A^\alpha_\mu\leftrightarrow\exp\biggr[ i \frac{\xi\cdot Q}{R}\, \mathbb{X}\biggr] \qquad , \qquad \tilde{A}^\alpha_\mu\leftrightarrow\exp\bigr[i (\xi\cdot Q) N^2 R\,\tilde{\mathbb{X}}\bigr] \,,
\end{equation}
respectively, and we display explicitly the radius dependence on the gauge bosons in order to stress the difference in their masses:
\begin{equation}
	M^2[A_\mu^\alpha] = \Bigr(\frac{\xi\cdot Q}{R}\Bigr)^2 \qquad , \qquad M^2[\tilde{A}_\mu^\alpha] = (\xi\cdot Q)^2 (N^2R)^2 \,.
\end{equation}
There are three cases of interest, corresponding to the possible values of the Scherk-Schwarz radius\footnote{Recall that the Scherk-Schwarz radius $R$ differs from the radius in the freely-acting orbifold picture by a rescaling, $R_{\textrm{orb}}=NR$.} $R$:
\begin{itemize}
	\item $R\gg 1/N$ : The gauge bosons $\tilde{A}$ carrying non-trivial winding charges decouple, as they acquire a mass much larger than the mass of the gauge bosons $A$ carrying momentum charges. In this case, the relevant algebra of gaugings involving $A$ is effectively truncated to the one in the case of pure momentum shift in eq. \eqref{AlgMomentumShift}.
	\item $R\ll 1/N$ : In this case, it is the gauge bosons $A$ carrying non-trivial momentum charge which decouple and the relevant algebra of gaugings involving $\tilde{A}$ is effectively the one corresponding to pure winding shift in eq. \eqref{AlgWindingShift}.
	\item $R= 1/N$ : In this region, both the gauge bosons carrying momentum $A$ as well as those carrying winding charge $\tilde{A}$ acquire the same mass. In this case, the gauge algebra at the level of the effective supergravity theory is enhanced, involving generators $T_\alpha$ and $\tilde{T}_\alpha$ (resp. $\mathcal{Z}_I,\tilde{\mathcal{Z}}_I, \mathcal{X}^I, \tilde{\mathcal{X}}^I$):
\begin{equation}
\begin{split}
&[\mathcal{Z}_\mathbb{X},\mathcal{Z}_I]=\tfrac{1}{2}(F_L+F_R)_I{}^J \mathcal{Z}_J+\tfrac{1}{2}(F_L-F_R)_{IJ}\mathcal{X}^J\\
&[\mathcal{Z}_\mathbb{X},\mathcal{X}^I]=-\tfrac{1}{2}(F_L+F_R)_J{}^I \mathcal{X}^J+\tfrac{1}{2}(F_L-F_R)^{IJ}\mathcal{Z}_J\\
&[\mathcal{X}^\mathbb{X},\tilde{\mathcal{Z}}_I]=\tfrac{1}{2}(F_L+F_R)_I{}^J \tilde{\mathcal{Z}}_J+\tfrac{1}{2}(F_L-F_R)_{IJ}\tilde{\mathcal{X}}^J\\
&[\mathcal{X}^\mathbb{X},\tilde{\mathcal{X}}^I]=-\tfrac{1}{2}(F_L+F_R)_J{}^I \tilde{\mathcal{X}}^J+\tfrac{1}{2}(F_L-F_R)^{IJ}\tilde{\mathcal{Z}}_J\\
\end{split}\,.\label{AlgMomentumWindingShift}
\end{equation}
Aside from this effective doubling of the generators, there is no gauge-symmetry enhancement with respect to the base, provided that the orbifold action preserves at least one supersymmetry from the left- and one from the right- moving sector. In cases when the (global) $N=2$ SCFT on the worldsheet is broken down to $N=1$ (either in the left- or right- movers), the $U(1)$ associated to the base $S^1$ can be enhanced to a non-abelian gauge symmetry at special values of the radius and the structure of the algebra in eq. \eqref{AlgMomentumWindingShift} may drastically change.

\end{itemize}


\subsection{$\mathbb{Z}_N\times\mathbb{Z}_M$ orbifolds with momentum and winding shift}\label{SecZnZmAlg}

We finally examine a further possibility, first proposed in \cite{Dabholkar:2005ve} in the context of non-geometric backgrounds,  involving freely-acting  (a)symmetric $\mathbb{Z}_N\times\mathbb{Z}_M$ orbifolds, where the first factor generates an order-$N$ momentum shift and the second factor an order-$M$ winding shift along the base $S^1$. Here, we will derive the general gauge algebra, whereas in Section \ref{ExampleSection} we will provide an explicit example of a consistent string vacuum of this type, based on an asymmetric $\mathbb{Z}_4\times\mathbb{Z}_2$ orbifold.

The action of the $\mathbb{Z}_N\times\mathbb{Z}_M$ orbifold can be represented in the following way:
\begin{equation}
	\begin{split}
	\mathcal{G} = e^{2\pi (F_L+F_R)}\, \delta_{1,0} \\
	\mathcal{G}' = e^{2\pi (F_L'+F_R')}\, \delta_{0,1}' \\
	\end{split}\,,
\end{equation}
where $F_L,F_R$ are the generators of order-$N$ rotations in the left- and right- moving coordinates of the fiber, associated to $\mathbb{Z}_N$, and $\delta_{1,0}$ is an order-$N$ momentum shift along the base $S^1$. Similarly, $F_L',F_R'$ are the order-$M$ rotation generators associated to $\mathbb{Z}_M$ and $\delta_{0,1}'$ is an order-$M$ winding shift in the base. The boundary conditions of the fermions can again be read from their contribution to the partition function:
\begin{equation}
	\prod_{i-\textrm{left}}\vartheta\biggr[{\alpha_i-2\xi_L^i H-2{\xi_L^i}' H' \atop \beta_i-2\xi_L^i G-2{\xi_L^j}' G'}\biggr] \ \prod_{j-\textrm{right}}\bar\vartheta\biggr[{\alpha_j-2\xi_R^j H -2{\xi_R^i}' H' \atop \beta_j-2\xi_R^j G-2{\xi_R^j}' G'}\biggr]\,.
\end{equation}
We again focus on the invariant states corresponding to gauge bosons, coming from the untwisted sector $H=H'=0$. The projection onto invariant states is enforced upon summation over $G\in\mathbb{Z}_N$ and $G'\in\mathbb{Z}_M$ and yields the constraints:
\begin{equation}
	\begin{split}
		m &=  N\xi\cdot Q \ (\textrm{mod}\ N) \\
		n &=  M\xi'\cdot Q \ (\textrm{mod}\ M) \\
	\end{split}\,.
\end{equation}
where $\xi\cdot Q\equiv \xi_L\cdot Q_L-\xi_R\cdot Q_R$ and $\xi'\cdot Q\equiv \xi_L'\cdot Q_L-\xi_R'\cdot Q_R$. The solution of the constraints then implies that the gauge bosons  acquire an internal part:
\begin{equation}
	A^\alpha_\mu\leftrightarrow \exp\biggr[i\frac{\xi\cdot Q}{R}\mathbb{X}+i(\xi'\cdot Q)\,NMR\,\tilde{\mathbb{X}}\biggr]\,,
\end{equation}
which precisely coincides with the ansatz in eq. \eqref{momwindreduction}, conjectured in \cite{Dabholkar:2005ve}.

The gauge bosons have their lowest mass at the self-dual (Scherk-Schwarz) radius $R=(NM)^{-1/2}$ and the algebra of gaugings takes the form:
\begin{equation}
\begin{split}
&[\mathcal{Z}_\mathbb{X},\mathcal{Z}_I]=\tfrac{1}{2}(F_L+F_R)_I{}^J \mathcal{Z}_J+\tfrac{1}{2}(F_L-F_R)_{IJ}\mathcal{X}^J\\
&[\mathcal{Z}_\mathbb{X},\mathcal{X}^I]=-\tfrac{1}{2}(F_L+F_R)_J{}^I \mathcal{X}^J+\tfrac{1}{2}(F_L-F_R)^{IJ}\mathcal{Z}_J\\
&[\mathcal{X}^\mathbb{X},\mathcal{Z}_I]=\tfrac{1}{2}(F_L'+F_R')_I{}^J \mathcal{Z}_J+\tfrac{1}{2}(F_L'-F_R')_{IJ}\mathcal{X}^J\\
&[\mathcal{X}^\mathbb{X},\mathcal{X}^I]=-\tfrac{1}{2}(F_L'+F_R')_J{}^I \mathcal{X}^J+\tfrac{1}{2}(F_L'-F_R')^{IJ}\mathcal{Z}_J\\
\end{split}\,.
\end{equation}
For a generic consistent choice of the flux matrices $F_L,F_R,F_L',F_R'$, the string background is ``truly'' non-geometric and contains at the same time $\omega,\tilde\omega,Q,\tilde{Q},H$ and $R$ fluxes, as can be verified by comparing with eqs. \eqref{generalgauging1}-\eqref{generalgauging3}.


\section{Orbifolds with Q/R-flux via T-duality}\label{SecTdualityOrb}

Having discussed the general framework which allows one to identify the fluxes through the gauging of the effective supergravity theory corresponding to (generically) asymmetric freely-acting orbifolds, we are now ready to study the fluxes appearing in various explicit examples. In this section, we will follow a chain of T-dualities which take us from the geometric flux, to the Q-flux and, finally, to the R-flux frame. The analysis based on freely-acting orbifolds has the advantage that, at each stage, one is dealing with an exactly solvable CFT, and the T-duality transformations can be followed exactly at the full string level. The main result is that an asymmetric action on the fiber (which parallels the asymmetric monodromy in the formalism of twisted reductions) is characterized by the presence of Q-flux, whereas an additional asymmetric action on the base gives rise to R-flux.


\subsection{General setup}\label{momShift}

Consider a compactification of Type II string theory on $(T^5\times S^1)/\mathbb{Z}_4$.
The freely-acting orbifold we will consider is the Type II analogue of the permutation orbifold originally studied in \cite{Kiritsis:1997ca} and is generated by:
\begin{equation}
	\mathcal{G} = e^{2\pi (F_L+ F_R)}\, \delta_{1,0} \,,
\end{equation}
where $\delta_{1,0}$ is an order-$4$ momentum shift and $F_L,F_R$ are the generators of rotations acting on the left- and right- moving coordinates, respectively. This class of orbifolds was later generalized to asymmetric versions in \cite{Condeescu:2012sp} and we will adopt the notation of the latter. For the purposes of the discussion, we will factorize the fiber $T^5$ as $T^4\times {S^1}'$, with ${S^1}'$ being a spectator circle, so that the orbifold acts symmetrically on the left- and right- moving  $T^4\times S^1$ coordinates:
\begin{equation}
	\begin{split}
	& Z^1 \rightarrow i Z^1\,\\
	& Z^2 \rightarrow -i Z^2 \, \\
\end{split}
\ ,\qquad
\begin{split}
	& Z^3 \rightarrow  i Z^3\, \\
	& Z^4 \rightarrow -i Z^4 \, \\
\end{split}
\ , \qquad
\begin{split}
	& \mathbb{X}\rightarrow \mathbb{X} + \frac{\pi}{2}
	\end{split}\,,\label{Z4action}
\end{equation}
where $Z^{1,2,3,4}$ are complex coordinates on $T^4$, related via complex conjugation $Z^2=\bar Z^1$ and $Z^4=\bar Z^3$.
On the other hand, $\mathbb{X}$ is the coordinate of the base, where the orbifold acts as a shift. The action on their fermionic superpartners is similar. The structure of the orbifold on the internal $T^4$ space is that of a K3 orbifold and, hence, preserves half of the left- and right- moving supersymmetries, giving rise to an $\mathcal{N}_4=4$ theory.

The partition function of the theory can be written in the Scherk-Schwarz formalism as a fibration:
\begin{equation}
\begin{split}
Z&=\sum_{m,n\in\mathbb{Z}}\frac{1}{2}\sum_{a,b=0,1}(-1)^{a+b}\frac{\vartheta^2\left[{a \atop b}\right] \vartheta\left[{a-h \atop b-g}\right]\vartheta\left[{a+h \atop b+g}\right]}{\eta^{12}}\frac{1}{2}\sum_{\bar a, \bar b=0,1}\frac{\bar\vartheta^2\left[{\bar a \atop \bar b}\right]\bar\vartheta\left[{\bar a-h \atop \bar b-g}\right]\bar\vartheta\left[{\bar a+h \atop \bar b+g}\right]}{\bar\eta^{12}}\\
&\times \Gamma_{(4,4)}\left[h \atop g\right]R\,\exp \left[-\frac{\pi R^2}{\tau_2}|\tilde m+\tau n|^2\right]\Gamma_{(1,1)}(R')\\
\end{split}\,,
\end{equation}
with $(h,g)=\frac{1}{2}(n,\tilde m)$. Here, $n,\tilde m\in\mathbb{Z}$ are the winding numbers parametrizing the wrapping of the worldsheet torus around $S^1$. Furthermore, the spectator circle ${S^1}'$ of radius $R'$ contributes the $\Gamma_{(1,1)}(R')$ spectator lattice. The lattice sum $\Gamma_{(4,4)}$ associated to the directions of the fiber $T^4$ on which the $\mathbb{Z}_4$ orbifold acts non-trivially is given at the fermionic point by:
\begin{equation}
\Gamma_{(4,4)}\left[h\atop g\right]=\frac{1}{2}\sum_{\gamma, \delta=0,1}\ \biggr|\,\vartheta\left[{\gamma -h \atop \delta-g}\right]\vartheta\left[{\gamma \atop \delta}\right]\vartheta\left[{\gamma +h \atop \delta+g}\right]\vartheta\left[{\gamma -2h \atop \delta-2g}\right]\,\biggr|^2
\,.
\end{equation}

Following \cite{Condeescu:2012sp},  the orbifold action can be related to the monodromy matrix of twisted reductions  by expressing the permutation action on the left-moving part of the fiber $X^I$:
\begin{equation}
	{(P_L)^I}_J = \left(\begin{array}{c r c r}
				0 & -1 & 0 & 0 \\
				1 & 0 & 0 & 0\\
				0 & 0 & 0 & -1 \\
				0 & 0 & 1 & 0 \\
				\end{array}\right) \equiv {(e^{2\pi F_L})^I}_J \,,
\end{equation}
 in terms of the left-moving \emph{flux matrix}  :
\begin{equation}
		{(F_L)^I}_J = \left(\begin{array}{c c c c}
				0 & -\frac{1}{4} & 0 & 0 \\
				\frac{1}{4} & 0 & 0 & 0\\
				0 & 0 & 0 & -\frac{1}{4} \\
				0 & 0 & \frac{1}{4} & 0 \\
				\end{array}\right) \, .\label{fluxmatrix}
\end{equation}
Since we are dealing with a symmetric orbifold, the action on the right-movers is identical, ${(F_R)^I}_J={(F_L)^I}_J\equiv {F^I}_J$. The flux matrices correspond precisely to the left- and right- moving data encoded into the monodromy matrix $M$ appearing in the twisted reduction \eqref{twisted} and, hence, they define a flat gauging of supergravity based on the algebra:
\begin{equation}
\begin{split}
&[\mathcal{Z}_\mathbb{X},\mathcal{Z}_I]= F_I{}^J \mathcal{Z}_J \\
&[\mathcal{Z}_\mathbb{X},\mathcal{X}^I]=-F_J{}^I \mathcal{X}^J\\
\end{split}\,.
\end{equation}
Comparing with the general form of the algebra of gaugings in eqs. \eqref{generalgauging1}-\eqref{generalgauging3}, one may verify that the model indeed corresponds to a geometric fibration and is, hence, an example of an $\omega$-flux background, with the identification $\omega^I_{\mathbb{X}J}=\tilde{\omega}^I_{\mathbb{X}J}=-{F^I}_J$.


\subsection{T-duality in the fiber}\label{SecTdualityFiber}

We will now perform a number of T-dualities, in order to investigate the manifestation of $Q$-flux and $R$-flux from the worldsheet perspective. It should be noted that  the naive application of Buscher rules is not valid for a generic compact manifold and the T-duality transformations typically need to be $\alpha'$-corrected order by order \cite{Serone:2003sv,Serone:2005ge}. We will not be concerned with such issues here, since our construction is based on an exactly solvable CFT and the T-dualities we perform can be exactly realized at the $\sigma$-model level.

The first step will be to perform a T-duality in the fiber directions. Let us first express the $T^4$ coordinates $Z^{1,2,3,4}$ in a  real basis $X^{1,2,3,4}$ :
\begin{equation}
	\begin{split}
		& Z^1_{L,R}= \tfrac{1}{2}(X^1+iX^2)_{L,R} \\
		& Z^2_{L,R}= \tfrac{1}{2}(X^1-iX^2)_{L,R} \\
	\end{split}
	\ \  ,\qquad
	\begin{split}
		& Z^3_{L,R}= \tfrac{1}{2}(X^3+iX^4)_{L,R} \\
		& Z^4_{L,R}= \tfrac{1}{2}(X^3-iX^4)_{L,R} \\
	\end{split}
\,. \label{T4coords1}
\end{equation}
The monodromy of $T^4$, as the base coordinate closes a full circle $\mathbb{X}\rightarrow \mathbb{X}+2\pi R$, acts as a permutation $X^1\rightarrow -X^2$, $X^2\rightarrow X^1$, $X^3\leftrightarrow -X^4$, $X^4\rightarrow X^3$, which reproduces precisely the orbifold action in eq. \eqref{Z4action}. Now let us perform a simultaneous T-duality on the coordinates $X^2, X^4 \in T^4$  of the fiber. Since the sigma model is flat toroidal, the T-duality action reads simply $X^I_L\rightarrow X^I_L$, $X^I_R\rightarrow - X^I_R$ and eq. \eqref{T4coords1} becomes in the dual theory:
\begin{equation}
	\begin{split}
		& Z^1_L= \tfrac{1}{2}(X^1+iX^2)_L \\
		& Z^2_L= \tfrac{1}{2}(X^1-iX^2)_L \\
		& Z^3_L= \tfrac{1}{2}(X^3+iX^4)_L \\
		& Z^4_L= \tfrac{1}{2}(X^3-iX^4)_L \\
	\end{split}
\quad,\quad
		\begin{split}
		& Z^1_R= \tfrac{1}{2}(X^1-iX^2)_R \\
		& Z^2_R= \tfrac{1}{2}(X^1+iX^2)_R \\
		& Z^3_R= \tfrac{1}{2}(X^3-iX^4)_R \\
		& Z^4_R= \tfrac{1}{2}(X^3+iX^4)_R\\
	\end{split}\quad . \label{T4coords2}
\end{equation}
Hence, in the T-dual theory, the monodromy becomes asymmetric between the left- and right- movers:
\begin{equation}
	\begin{split}
		& Z^1_L \rightarrow +i Z^1_L\\
		& Z^2_L \rightarrow -i Z^2_L\\
		& Z^3_L \rightarrow +i Z^3_L\\
		& Z^4_L \rightarrow -i Z^4_L\\
	\end{split}
 \quad,\quad
		\begin{split}
		& Z^1_R \rightarrow -i Z^1_R\\
		& Z^2_R \rightarrow +i Z^2_R\\
		& Z^3_R \rightarrow -i Z^3_R\\
		& Z^4_R \rightarrow +i Z^4_R\\
	\end{split}\quad .
\end{equation}
Of course, the partition function remains unchanged in the T-dual theory. The spacetime interpretation, however, is now different. Due to the asymmetric monodromy ${(F_L)^I}_J=-{(F_R)^I}_J\equiv {F^I}_J$, one expects the model to contain $Q$-flux. This can be best seen by inspecting the algebra of gaugings:
\begin{equation}
\begin{split}
&[\mathcal{Z}_\mathbb{X},\mathcal{Z}_I]=F_{IJ}\mathcal{X}^J\\
&[\mathcal{Z}_\mathbb{X},\mathcal{X}^I]=F^{IJ}\mathcal{Z}_J\\
\end{split}\,,
\end{equation}
from which one may infer the presence of a combination of both $H$- and $Q$- flux, $H_{\mathbb{X}IJ}=F_{IJ}$, $Q_\mathbb{X}^{IJ}=-F^{IJ}$.


\subsection{T-duality in the base}\label{windShift}

In order to investigate the worldsheet interpretation of the R-flux, we will now perform a T-duality in the base $S^1$. Of course, for a generic twisted torus, such an operation is not allowed because the moduli of the fiber depend explicitly on the base coordinate $\mathbb{X}$  and the $U(1)$-shift along the base is not an isometry of the background. On the other hand, in the special case where the fibration is realized at the orbifold point, the shift symmetry along the base  is still present, since the twist of the fiber depends on the winding around the base $S^1$ only through non-trivial boundary conditions of the worldsheet fields. This allows one to perform the T-duality in an exact way at the level of the $\sigma$-model.

We start from the flat $\sigma$-model action, ignoring the dilaton:
\begin{equation}
	S = \frac{1}{4\pi}\int d^2 \sigma \ (G_{IJ}\partial X^I \bar\partial X^J + R^2 \partial \mathbb{X} \bar\partial\mathbb{X})+\ldots\,,
\end{equation}
where we only display the relevant bosonic fields. For the symmetric $\mathbb{Z}_4$-orbifold of the previous section the radii of the $T^4$ can be taken to lie at the fermionic point, $G_{IJ}=\frac{1}{2}\delta_{IJ}$. In the full path integral, one has to supplement this with a set of non-trivial boundary conditions. To illustrate these boundary condition assignments, it is more convenient to rewrite the partition function in the orbifold representation, by redefining the windings as $n\rightarrow 4n+H$, $\tilde{m}\rightarrow 4\tilde{m}+G$ with $n,\tilde{m}\in\mathbb{Z}$, $H,G\in\mathbb{Z}_4$ and using the periodicities of the Jacobi $\vartheta$-functions to decouple the winding dependence of the latter:
\begin{equation}
\begin{split}
Z&=\frac{1}{4}\sum_{H,G\in\mathbb{Z}_4}\frac{1}{2}\sum_{a,b=0,1}(-1)^{a+b}\frac{\vartheta^2\left[{a \atop b}\right] \vartheta\left[{a-H \atop b-G}\right]\vartheta\left[{a+H \atop b+G}\right]}{\eta^{12}}\frac{1}{2}\sum_{\bar a, \bar b=0,1}\frac{\bar\vartheta^2\left[{\bar a \atop \bar b}\right]\bar\vartheta\left[{\bar a-H \atop \bar b-G}\right]\bar\vartheta\left[{\bar a+H \atop \bar b+G}\right]}{\bar\eta^{12}}\\
&\times \Gamma_{(4,4)}\left[H \atop G\right]4R\sum_{\tilde{m},n\in\mathbb{Z}}\exp \left[-\frac{\pi (4R)^2}{\tau_2}\Bigr|\tilde m+\frac{G}{4}+\tau\Bigr( n+\frac{H}{4}\Bigr)\Bigr|^2\right]\Gamma_{(1,1)}(R')\\
\end{split}\,.\label{OriginalPartition}
\end{equation}
This representation\footnote{Notice the effective rescaling of the radius $R\rightarrow 4R$, which takes place when one goes from the Scherk-Schwarz representation to the orbifold picture.} of the partition function illustrates the boundary conditions of the various worldsheet fields. The path integral on the worldsheet torus for  the base coordinate $\mathbb{X}$ is performed by splitting the latter into a classical (instantonic) part, encoding the boundary conditions, and a quantum part $\mathbb{X}_{\textrm{quant}}$ with trivial boundary conditions which gives rise to the string oscillator contribution:
\begin{equation}
	\mathbb{X}(\sigma^1,\sigma^2) = 2\pi\Bigr[ (n+\tfrac{H}{4})\sigma^1+(\tilde m+\tfrac{G}{4})\sigma^2\Bigr] +\mathbb{X}_{\textrm{quant}}\,. \label{ClassPartX}
\end{equation}
The flat worldsheet $\sigma$-model has a local isometry under translations in the base direction $\mathbb{X}$ with conserved Noether currents  $J=(4R)^2 \partial\mathbb{X}$, $\bar{J}=(4R)^2 \bar\partial\mathbb{X}$. This isometry can be gauged \cite{Buscher:1987sk} by introducing a term $\int(A\bar{J}+\bar{A}J)$ in the worldsheet action, coupling the currents to appropriate gauge fields $A, \bar{A}$. However, the new term still varies under the gauge transformation and gauge invariance can be ensured by adding a quadratic term $\int (4R)^2 A\bar{A}$ in the gauge fields, together with a term $\int (A\bar\partial\mathbb{Y}-\bar{A}\partial\mathbb{Y})$, involving an additional real scalar $\mathbb{Y}$. The gauged action then reads:
\begin{equation}
	\begin{split}
	S' = \frac{1}{4\pi}\int & d^2\sigma\, \Bigr\{  G_{IJ}\partial X^I\bar\partial X^J+(4R)^2\partial\mathbb{X}\bar\partial\mathbb{X} \\
	&+ \Bigr[(4R)^2\partial\mathbb{X}-\partial\mathbb{Y}\Bigr]\bar{A}+\Bigr[(4R)^2\bar\partial\mathbb{X}+\bar\partial\mathbb{Y}\Bigr]A+(4R)^2 A\bar{A} \,\Bigr\}
	\end{split}\,.
\end{equation}
The quantum equivalence of the gauged action $S'$ to the original one can be seen by carefully integrating out $\mathbb{Y}$ on the topology of the worldsheet torus which, in turn, forces  $A$ to be a pure gauge \cite{Buscher:1987sk}. The dual action, on the other hand, is obtained by instead integrating out the gauge fields:
\begin{equation}
	\begin{split}
	A = -\partial\mathbb{X}+(4R)^{-2}\partial\mathbb{Y} \ \ , \qquad \bar{A} = - \bar\partial\mathbb{X} - (4R)^{-2}\bar\partial\mathbb{Y}
	\end{split}\,.
\end{equation}
Substituting these into $S'$ then leads to the dual action $\tilde{S}$:
\begin{equation}
	\tilde{S} =\frac{1}{4\pi}\int d^2\sigma\,\bigr[ G_{IJ}\partial X^I\bar\partial X^J + (4R)^{-2}\partial\mathbb{Y}\bar\partial\mathbb{Y} + (\partial \mathbb{Y}\bar\partial\mathbb{X}-\partial\mathbb{X}\bar\partial\mathbb{Y})\bigr]\,,
\end{equation}
with $\mathbb{Y}$ being interpreted as the T-dual coordinate of the base. Notice that the original base coordinate $\mathbb{X}$ still appears in the dual action, through its coupling to $\mathbb{Y}$ via a total-derivative term:
\begin{equation}
	S_{\mathbb{XY}}=\frac{1}{4\pi}\int d^2\sigma\ (\partial \mathbb{Y}\bar\partial\mathbb{X}-\partial\mathbb{X}\bar\partial\mathbb{Y})\,.\label{extraterm}
\end{equation}
On a worldsheet with the topology of a sphere this term would simply drop out, leading to the standard T-dual action. In the case of the worldsheet torus, however, the extra term in eq. \eqref{extraterm} does contribute through the classical part of $\mathbb{X},\mathbb{Y}$. Using the torus derivatives:
\begin{equation}
	\begin{split}
	\partial = \frac{i}{\sqrt{\tau_2}}( \bar\tau \partial_1-\partial_2 ) \ \ , \qquad
	\bar\partial = \frac{i}{\sqrt{\tau_2}}(-\tau\partial_1 + \partial_2)
	\end{split}\,,
\end{equation}
together with the classical part of $\mathbb{X}$, given in eq.\eqref{ClassPartX}, and that of $\mathbb{Y}$:
\begin{equation}
	\mathbb{Y}(\sigma^1,\sigma^2) = 2\pi\left( n'\sigma^1+\tilde m'\sigma^2\right) +\mathbb{Y}_{\textrm{quant}}\,,
\end{equation}
one obtains the non-trivial contribution to the partition function:
\begin{equation}
	\exp[-S_{\mathbb{XY}}] = \exp\Bigr[2i\pi\Bigr(\frac{\tilde{m}'H-n'G}{4}\Bigr)\Bigr]\,.\label{extraphase}
\end{equation}
Hence, the windings $n,\tilde{m}$ of the original base coordinate $\mathbb{X}$ drop out and, hence, the phase in eq. \eqref{extraphase}, together with the standard kinetic term of $\mathbb{Y}$ in the dual $\sigma$-model form the T-dual $S^1$ lattice:
\begin{equation}
	\Gamma_{(1,1)}\Bigr[{H \atop G}\Bigr]=\frac{(4R)^{-1}}{\sqrt{\tau_2}}\sum\limits_{\tilde{m}',n'\in\mathbb{Z}} \exp\Biggr[-\frac{\pi(4R)^{-2}}{\tau_2}\bigr|\tilde m'+\tau n'\bigr|^2+2i\pi\Bigr(\frac{\tilde{m}'H-n'G}{4}\Bigr)\Biggr]\,.
\end{equation}
Similarly, path integration over the remaining worldsheet bosons and their fermionic superpartners yields the remaining pieces of the partition function of the T-dual theory:
\begin{equation}
\begin{split}
Z&=\frac{1}{4}\sum_{H,G\in\mathbb{Z}_4}\frac{1}{2}\sum_{a,b=0,1}(-1)^{a+b}\frac{\vartheta^2\left[{a \atop b}\right] \vartheta\left[{a-H \atop b-G}\right]\vartheta\left[{a+H \atop b+G}\right]}{\eta^{12}}\frac{1}{2}\sum_{\bar a, \bar b=0,1}\frac{\bar\vartheta^2\left[{\bar a \atop \bar b}\right]\bar\vartheta\left[{\bar a-H \atop \bar b-G}\right]\bar\vartheta\left[{\bar a+H \atop \bar b+G}\right]}{\bar\eta^{12}}\\
&\times \Gamma_{(4,4)}\left[H \atop G\right]\frac{1}{4R}\sum_{\tilde{m}',n'\in\mathbb{Z}}\exp\Biggr[-\frac{\pi(4R)^{-2}}{\tau_2}\bigr|\tilde m'+\tau n'\bigr|^2+2i\pi\Bigr(\frac{\tilde{m}'H-n'G}{4}\Bigr)\Biggr]\Gamma_{(1,1)}(R')\\
\end{split}\,.
\end{equation}
Notice that the above representation of the partition function could have been directly obtained from eq. \eqref{OriginalPartition} by Poisson-resumming first in the $\tilde{m}$- and then in the $n$- winding. The interpretation of the T-dual theory is most clear in the Hamiltonian representation of the $S^1$-lattice:
\begin{equation}
	\Gamma_{(1,1)}\Bigr[{H \atop G}\Bigr]=\sum\limits_{m,n\in\mathbb{Z}} q^{\frac{1}{2}P_L^2}\ \bar{q}^{\frac{1}{2}P_R^2}\ e^{i\pi  n G/2}\,,
\end{equation}
with
\begin{equation}
	P_{L,R} =\frac{1}{\sqrt{2}}\left(\frac{m+\frac{H}{4}}{4R}\pm 4R n\right)\,,
\end{equation}
from which one recognizes a winding shift (see Section \ref{SecGeneralShifts}). Hence, the effect of T-duality in the base $S^1$ is to turn the original freely-acting orbifold with momentum shift along $S^1$ into a freely-acting orbifold with identical action on the fiber, but with winding shift along the base. The winding shift involves an asymmetric action of the orbifold on the base and, in the case when the action on the fiber is also asymmetric, corresponds to the $R$-flux frame. Let us make this identification explicitly in the two relevant cases:
\begin{itemize}
	\item Starting from an orbifold with momentum shift, which acts symmetrically on the fiber ($F_L=F_R\equiv F$), and performing a  T-duality in the base $S^1$,  yields a non-geometric background corresponding to the gauge algebra:
		\begin{equation}
\begin{split}
&[\mathcal{X}^\mathbb{X},\mathcal{Z}_I]=F_I{}^J \mathcal{Z}_J\\
&[\mathcal{X}^\mathbb{X},\mathcal{X}^I]=-F_J{}^I \mathcal{X}^J\\
\end{split}\,.
\end{equation}
The resulting gauging of supergravity corresponds to  $Q$-flux only, with the identification $Q^{\mathbb{X}J}_{I}=-\tilde{Q}^{\mathbb{X}I}_{J}={F_I}^J$.

	\item If, on the other hand, we first perform a T-duality in the fiber directions, so that the action of the orbifold is asymmetric on the fiber ($F_L=-F_R\equiv F$), then the subsequent T-duality in the base $S^1$ yields:
	\begin{equation}
\begin{split}
&[\mathcal{X}^\mathbb{X},\mathcal{Z}_I]=F_{IJ}\mathcal{X}^J\\
&[\mathcal{X}^\mathbb{X},\mathcal{X}^I]=F^{IJ}\mathcal{Z}_J\\
\end{split}\,.
\end{equation}
This case is characterized by a combination of both $\omega$- and $R$-flux, with the identifications $\tilde{\omega}^\mathbb{X}_{IJ}=F_{IJ}$, $R^{\mathbb{X}IJ}=F^{IJ}$, respectively.

\end{itemize}



\section{Asymmetric orbifolds without symmetric T-duals}\label{ExampleSection}

In the previous section, we started from the simple example of a freely-acting, symmetric orbifold and studied a chain of T-dualities which effectively turned the action on the fiber and on the base into an asymmetric one. By studying the algebra of gaugings of the corresponding effective supergravity theory, we were able to identify the $\omega, H, Q$ and $R$ fluxes present in each case. These string backgrounds lie inside the same conjugacy class of $O(D,D;\mathbb{Z})$ as the geometric fibration and, hence, constitute a description of the same (initially geometric) theory in different duality frames. On the other hand, at the level of the supergravity theory there is a plethora of possible gaugings which have no geometric origin. These are theories which lie outside the geometric conjugacy class of $O(D,D;\mathbb{Z})$ and, hence, their underlying string theoretic description is non-trivially and inherently non-geometric. In this section, we will study such ``truly'' asymmetric constructions at the string level in terms of explicit examples.


\subsection{Inherently asymmetric $\mathbb{Z}_4$ orbifold}\label{SecExtremeAsymZ4}

Consider now the freely-acting $\mathbb{Z}_4$ orbifold:
\begin{equation}
	\mathcal{G} = e^{2\pi F_L}\, \delta_{a,b} \,,
\end{equation}
where the orbifold acts as a rotation $F_L$ only on the left-moving coordinates of the fiber, accompanied with a generic shift $\delta_{a,b}$ along the base $S^1$. This is an ``extremely'' asymmetric analogue of the $\mathbb{Z}_4$ orbifolds discussed in the previous section. The asymmetry of the fiber cannot be removed by any action of the T-duality group and, in the case $a=b=1$ corresponding to a combined momentum and winding shift, the base also acquires an inherent stringy asymmetry.
The modular invariant partition function corresponding to the left-moving flux matrix $F_L$ given in eq. \eqref{fluxmatrix} is:
\begin{equation}
\begin{split}
Z&=\frac{1}{4}\sum_{H,G\in\mathbb{Z}_4}\frac{1}{2}\sum_{\alpha,\beta=0,1}(-1)^{\alpha+\beta}\frac{\vartheta^2\left[{\alpha \atop \beta}\right] \vartheta\left[{\alpha-H/2 \atop \beta-G/2}\right]\vartheta\left[{\alpha+H/2 \atop \beta+G/2}\right]}{\eta^{12}}\frac{1}{2}\sum_{\bar a, \bar b=0,1}\frac{\bar\vartheta^4\left[{\bar a \atop \bar b}\right]}{\bar\eta^{12}}\\
&\times \Gamma_{(4,4)}\left[H \atop G\right]\Gamma_{(1,1)}(R')~ e^{2\pi i[ ab HG/4- (HG-\delta H)]/4} \sum\limits_{m,n\in\mathbb{Z}} e^{2i\pi G  \frac{m a+n b}{4}} \, q^{\frac{1}{4}P_L^2}\,\bar{q}^{\frac{1}{4}P_R^2} \\
\end{split}\,,
\end{equation}
where the left- and right- moving momenta along the base are given by:
\begin{equation}
	P_{L,R} = \frac{m+b\frac{H}{4}}{4R}\pm 4R\left(n+a\tfrac{H}{4}\right) \,.
\end{equation}
The asymmetrically twisted $\Gamma_{(4,4)}$-lattice associated to the $T^4$ fiber, now reads (at the fermionic point):
\begin{equation}
\Gamma_{(4,4)}\left[H\atop G\right]=\frac{1}{2}\sum_{\gamma, \delta=0,1} \vartheta\left[{\gamma -H/2 \atop \delta-G/2}\right]\vartheta\left[{\gamma \atop \delta}\right]\vartheta\left[{\gamma +H/2 \atop \delta+G/2}\right]\vartheta\left[{\gamma -H \atop \delta-G}\right]\ \bar\vartheta\left[{\gamma\atop\delta}\right]^4 \,.
\end{equation}
There are three cases of interest, corresponding to pure momentum shift $\delta_{1,0}$, pure winding shift $\delta_{0,1}$ and a combined momentum and winding shift $\delta_{1,1}$. It will be instructive to display explicitly the algebra of gaugings in each of these cases.

\begin{itemize}

\item Orbifold with pure momentum shift $\delta_{1,0}$:
\begin{equation}
\begin{split}
&[\mathcal{Z}_\mathbb{X},\mathcal{Z}_I]=\tfrac{1}{2}F_I{}^J \mathcal{Z}_J+\tfrac{1}{2} F_{IJ}\mathcal{X}^J\\
&[\mathcal{Z}_\mathbb{X},\mathcal{X}^I]=-\tfrac{1}{2}F_J{}^I \mathcal{X}^J+\tfrac{1}{2}F^{IJ}\mathcal{Z}_J\\
\end{split}\,.
\end{equation}
Hence, the theory contains a combination of $\omega, H$ and $Q$ flux, given by: $\omega^J_{\mathbb{X}I}=\tilde\omega^J_{\mathbb{X}I}=\frac{1}{2}{F_I}^J$, $H_{\mathbb{X}IJ}=\frac{1}{2}F_{IJ}$ and $Q_\mathbb{X}^{IJ}=\frac{1}{2}F^{IJ}$.

\item Orbifold with pure winding shift $\delta_{0,1}$:
\begin{equation}
\begin{split}
&[\mathcal{X}^\mathbb{X},\mathcal{Z}_I]=\tfrac{1}{2}F_I{}^J \mathcal{Z}_J+\tfrac{1}{2}F_{IJ}\mathcal{X}^J\\
&[\mathcal{X}^\mathbb{X},\mathcal{X}^I]=-\tfrac{1}{2}F_J{}^I \mathcal{X}^J+\tfrac{1}{2}F^{IJ}\mathcal{Z}_J\\
\end{split}\,.
\end{equation}
This corresponds to the T-dual of the previous case with respect to the base and contains $\omega, Q$ and $R$ flux. The precise identification is given by: $\tilde\omega^\mathbb{X}_{IJ}=\frac{1}{2}F_{IJ}$, $Q^{\mathbb{X}J}_{I}=-\tilde{Q}^{\mathbb{X}J}_{I}=\frac{1}{2}{F_I}^J$ and $R^{\mathbb{X}IJ}=\frac{1}{2}F^{IJ}$.

\item Orbifold with momentum and winding shift $\delta_{1,1}$:
\begin{equation}
\begin{split}
&[\mathcal{Z}_\mathbb{X},\mathcal{Z}_I]=\tfrac{1}{2}F_I{}^J \mathcal{Z}_J+\tfrac{1}{2}F_{IJ}\mathcal{X}^J\\
&[\mathcal{Z}_\mathbb{X},\mathcal{X}^I]=-\tfrac{1}{2}F_J{}^I \mathcal{X}^J+\tfrac{1}{2}F^{IJ}\mathcal{Z}_J\\
&[\mathcal{X}^\mathbb{X},\tilde{\mathcal{Z}}_I]=\tfrac{1}{2}F_I{}^J \tilde{\mathcal{Z}}_J+\tfrac{1}{2}F_{IJ}\tilde{\mathcal{X}}^J\\
&[\mathcal{X}^\mathbb{X},\tilde{\mathcal{X}}^I]=-\tfrac{1}{2}F_J{}^I \tilde{\mathcal{X}}^J+\tfrac{1}{2}F^{IJ}\tilde{\mathcal{Z}}_J\\
\end{split}\,.
\end{equation}
As discussed in Section \ref{SecAlgMomentumWindingShift}, the enhancement of the algebra is an inherently stringy phenomenon with no geometric analogue and corresponds to a gauging involving simultaneously $\omega, H, Q$ and $R$ flux.

\end{itemize}

For simplicity we defined $F_L\equiv F$. Notice that,  due to the asymmetry of the rotation, the effective quantization of the rotation angles in the flux matrix $F$ (as it appears in the algebra) is in units of $1/(2N)$, unlike in the case of a symmetric action where the quantization is in units of $1/N$. This is a generic phenomenon arising in such ``extremely'' asymmetric constructions where the action of the orbifold rotation acts chirally  either only on the left- or right- moving fiber coordinates.


\subsection{$\mathbb{Z}_4\times\mathbb{Z}_2$ orbifold with momentum and winding shift}\label{Z2Z4orbCFT}

In this section we discuss a generalization of this class of orbifold backgrounds to freely-acting, asymmetric $\mathbb{Z}_4\times\mathbb{Z}_2$ orbifolds, with the $\mathbb{Z}_4$ factor acting on the left-moving fiber, accompanied with an order-4 momentum shift in the base and the $\mathbb{Z}_2$ factor acting on the right-moving fiber coordinates, together with an order-2 winding shift on the base:
\begin{equation}
	\begin{split}
	\mathcal{G} = e^{2\pi F_L}\, \delta_{1,0} \\
	\mathcal{G}' = e^{2\pi F_R'}\, \delta_{0,1}' \\
	\end{split}\,.
\end{equation}
The modular invariant partition function is given by:
\begin{equation}
\begin{split}
Z&=\frac{1}{4}\sum_{H,G\in\mathbb{Z}_4}\frac{1}{2}\sum_{H',G'\in\mathbb{Z}_2}\frac{1}{2}\sum_{\alpha,\beta=0,1}(-1)^{\alpha+\beta}\frac{\vartheta^2\left[{\alpha \atop \beta}\right] \vartheta\left[{\alpha-H/2 \atop \beta-G/2}\right]\vartheta\left[{\alpha+H/2 \atop \beta+G/2}\right]}{\eta^{12}}\\
&\times \frac{1}{2}\sum_{\bar\alpha,\bar\beta=0,1}(-1)^{\bar\alpha+\bar\beta}\frac{\bar\vartheta^2\left[{\bar\alpha \atop \bar\beta}\right] \bar\vartheta\left[{\bar\alpha-H'/2 \atop \bar\beta-G'/2}\right]\bar\vartheta\left[{\bar\alpha+H'/2 \atop \bar\beta+G'/2}\right]}{\bar\eta^{12}}\\
&\times \Gamma_{(4,4)}\left[H , H' \atop G , G'\right]\Gamma_{(1,1)}(R')~ e^{\pi i[\frac{HG-\delta H}{2}-\frac{H'G}{4}-H'G']} \sum\limits_{m,n\in\mathbb{Z}} e^{2i\pi [G m/4 - G'n/2]} \, q^{\frac{1}{4}P_L^2}\,\bar{q}^{\frac{1}{4}P_R^2} \\
\end{split}\,,
\end{equation}
where the left- and right- moving momenta along the base are given by:
\begin{equation}
	P_{L,R} = \frac{m-\frac{H'}{2}}{4R}\pm 4R\left(n+\tfrac{H}{4}\right) \,.
\end{equation}
The asymmetrically twisted $\Gamma_{(4,4)}$-lattice associated to the $T^4$ fiber, now reads (at the fermionic point):
\begin{equation}
\Gamma_{(4,4)}\left[H,H'\atop G,G'\right]=\frac{1}{2}\sum_{\gamma, \delta=0,1} \vartheta\left[{\gamma -H/2 \atop \delta-G/2}\right]\vartheta\left[{\gamma \atop \delta}\right]\vartheta\left[{\gamma +H/2 \atop \delta+G/2}\right]\vartheta\left[{\gamma -H \atop \delta-G}\right] \bar\vartheta\left[{\gamma\atop\delta}\right]^2 \bar\vartheta\left[{\gamma-H' \atop \delta-G'}\right]\bar\vartheta\left[{\gamma+H'\atop\delta+G'}\right] \,.
\end{equation}
The orbifold action on the fiber is parametrized through the flux matrices $F_L, F_R'$, where $F_L$ is the generator of the $\mathbb{Z}_4$ rotation on the left-movers given in eq. \eqref{fluxmatrix} and $F_R$ is given by:
\begin{equation}
		{(F_R')^I}_J = \left(\begin{array}{c c c c}
				0 & -\frac{1}{2} & 0 & 0 \\
				\frac{1}{2} & 0 & 0 & 0\\
				0 & 0 & 0 & -\frac{1}{2} \\
				0 & 0 & \frac{1}{2} & 0 \\
				\end{array}\right) \, .
\end{equation}
In this case, the resulting algebra of gaugings becomes:
\begin{equation}
\begin{split}
&[\mathcal{Z}_\mathbb{X},\mathcal{Z}_I]=\tfrac{1}{2}(F_L)_I{}^J \mathcal{Z}_J+\tfrac{1}{2}(F_L)_{IJ}\mathcal{X}^J\\
&[\mathcal{Z}_\mathbb{X},\mathcal{X}^I]=-\tfrac{1}{2}(F_L)_J{}^I \mathcal{X}^J+\tfrac{1}{2}(F_L)^{IJ}\mathcal{Z}_J\\
&[\mathcal{X}^\mathbb{X},\mathcal{Z}_I]=\tfrac{1}{2}(F_R')_I{}^J {\mathcal{Z}}_J+\tfrac{1}{2}(F_R')_{IJ}{\mathcal{X}}^J\\
&[\mathcal{X}^\mathbb{X},{\mathcal{X}}^I]=-\tfrac{1}{2}(F_R')_J{}^I {\mathcal{X}}^J+\tfrac{1}{2}(F_R')^{IJ}{\mathcal{Z}}_J\\
\end{split}\,,\label{Z2Z4algebra}
\end{equation}
and corresponds to ``true" $Q$- and $R$- backgrounds with non-trivial $\omega, \tilde\omega, H, Q, \tilde{Q}$ and $R$ fluxes. It is important to stress that the fluxes appearing in the above algebra are  effectively quantized in units of $1/8$ for $F_L$ and $1/4$ for $F_R'$, and this necessitates the distinction between $\omega$, $\tilde\omega$ and $Q$, $\tilde{Q}$.


\section{On the relation between asymmetric orbifolds and T-folds}\label{BackgroundDesc}

Besides deriving the (non-) geometric fluxes and the corresponding supergravity gaugings from the CFT operator algebra, the relation between (a)symmetric orbifolds and (non-) geometric fluxes
can be also seen from a slightly different but eventually equivalent perspective, namely by viewing the orbifolds as particular points in the parameter space of a T-fold compactification. The main idea is that T-folds are defined by non-constant background fields, which are patched together either by standard diffeomorphisms (geometric spaces) or, more generally, by stringy symmetries that correspond to discrete symmetry operations of the orbifold CFT.
Allowing for generic background parameters (moduli), the $\sigma$-model associated to the T-fold is not, in general, conformal and the equations of motion will receive an infinite tower of $\alpha'$-corrections. However, choosing the moduli in such a way that the symmetry transformations act as automorphisms on the background, \emph{i.e.} going to the (a)symmetric orbifold points of the T-fold moduli space, the correspond $\sigma$-model  becomes conformal. Hence, there is a renormalization group flow in the moduli space of the T-fold towards the directions of its orbifold points, where some (or all)
of the moduli parameters are fixed to specific values\footnote{In the interest of simplicity, we use in this discussion the example of T-folds. However, the arguments we present remain valid even in the case of ``generalized T-folds" where the description is inherently non-local, namely for $R$-backgrounds.}. It turns out that these fixed values of the moduli precisely correspond to the minima of the gauged supergravity potential, consistently with the fact that the gauging can be equivalently viewed as a compactification with (non-) geometric fluxes.

The $D=(d+d')$-dimensional backgrounds ${\cal Y}^{d+d'}$ under consideration can be conveniently described in a uniform manner: they take the form of a fibration of a $d$-dimensional  torus $T_f^d$  over a $d'$-dimensional base ${\cal B}^{d'}$ in the remaining directions:
\begin{equation}\label{fibration}
T^d_{f}\,\hookrightarrow\, {\cal Y}^{d+d'}\,\hookrightarrow\, {\cal B}^{d'}\, .
\end{equation}
For instance, the base space ${\cal B}^{d'}$ can be taken to be a $d'$-dimensional torus $T_b^{d'}$.
We wish to consider fibrations that are determined by the $O(d,d)$ monodromy properties of the fiber torus $T_f^d$, when going around homologically non-trivial loops in the
base ${\cal B}^{d'}$.

Let us consider the case where, in the (a)symmetric orbifold language, the fibration is specified by the  $\mathbb{Z}_N^L\times\mathbb{Z}_M^R$ orbifold rotations $(\mathcal{M}_L$, $\mathcal{M}_R)$ acting on the left- and right-moving fiber ``coordinates" $X_L^I$ and $X_R^I$ ($I=1,\dots ,d$).
Namely, a shift in a base coordinate $\mathbb{X}\rightarrow \mathbb{X}+2\pi$ induces the following rotation on $X_L$ and $X_R$:
\begin{equation}
X_L\rightarrow {\cal M}_L X_L\, ,\quad X_R\rightarrow {\cal M}_R X_R\, .
\end{equation}
This defines a freely-acting orbifold space, since the fiber twist is always accompanied by a shift along the base coordinate $\mathbb{X}$. It corresponds to a geometric, symmetric orbifold when ${\cal M}_L={\cal M}_R$, whereas, otherwise, the underlying CFT corresponds to a non-geometric asymmetric orbifold.

Let us now see how the above orbifold picture is related to a T-fold description. In the T-fold picture, the background parameters corresponding to the metric and antisymmetric tensor field are non-constant
 functions,  varying over of the base coordinates $\mathbb{X}$: $g_{IJ}=g_{IJ}(\mathbb{X})$, $b_{IJ}=b_{IJ}(\mathbb{X})$. Consistency then requires, that the monodromy of the background around closed loops in the base (\emph{i.e.} $\mathbb{X}\rightarrow \mathbb{X}+2\pi$),
respects the stringy symmetries, which are given in terms of discrete $O(d,d)$ transformations.
To illustrate the latter, we combine the metric and the antisymmetric tensor of $T_f^d$ into a $d$-dimensional matrix as
\begin{equation}
{\cal E}_{IJ}(\mathbb{X}) =g_{IJ}(\mathbb{X})+b_{IJ}(\mathbb{X})\,,
\end{equation}
in terms of which, the $O(d,d)$ transformation of the background fields acts as
\begin{equation}
{\cal E}(\mathbb{X}+2 \pi) = {\cal M}_{O(d,d)} {\cal E}(\mathbb{X})=\Bigl(A{\cal E}(\mathbb{X})+B\Bigr)\Bigl(C{\cal E}(\mathbb{X})+D\Bigr)^{-1}\,.
\end{equation}
Here, ${\cal M}_{O(d,d)}$ is a  group element of $O(d,d)$  of the form
\begin{equation}\label{SOdmatrix}
{\cal M}_{O(d,d)}=\begin{pmatrix} A & B \\ C & D \end{pmatrix}\, ,
\end{equation}
where the $d$-dimensional matrices $A,B,C,D$ satisfy
\begin{equation}\label{constraints}
A^tC+C^tA=0\, ,\quad B^tD+D^tB=0\, ,\quad A^tD+C^tB=I\, .
\end{equation}
In order to match the orbifold symmetries, ${\cal M}_{SO(d,d)}$ must be identified with the rotation group elements $({\cal M}_L,{\cal M}_R)$ of the $\mathbb{Z}_N^L\times \mathbb{Z}_M^R$ orbifold.
Therefore, the identification between the T-fold and the freely-acting $\mathbb{Z}_N^L\times \mathbb{Z}_M^R$ orbifold is possible, provided that a faithful embedding of $\mathbb{Z}_N^L\times \mathbb{Z}_M^R$
into $O(d,d)$ can be found.
The general asymmetric case with ${\cal M}_L\neq {\cal M}_R$ corresponds to a non-geometric T-fold, and the combined $({\cal M}_L,{\cal M}_R)$ rotation forms a discrete element of the full
 $O(d,d)$ group; \emph{i.e.}  the group $\mathbb{Z}_N^L\times \mathbb{Z}_M^R$ is a discrete Abelian subgroup of $O(d,d)$. On the other hand, in the case of a symmetric orbifold, the symmetric rotation group with ${\cal M}_L={\cal M}_R$ is  a subgroup of only the diagonal $O(d)$. In this case the associated T-fold describes a geometric compactification.

We will  explicitly demonstrate this relationship by considering a three-dimensional fibration, with a one-dimensional
circle $S^1$ serving as the base and a two-dimensional torus $T^2$  as the fiber. A $T^2$ torus with $b$-field is parametrized by two complex scalars known as the complex structure $\tau={\frac{g_{12}}{g_{11}}}+i\ {\frac{V}{g_{11}}}$ and the complexified
K\"ahler class $\rho=-b_{12}+i\ V$, where $V$ denotes the volume of the two-torus.
The $O(2,2)$ group then decomposes as follows:
\begin{equation}
O(2,2)=SL(2)_\tau\times SL(2)_\rho\times \mathbb{Z}_2^{\tau\leftrightarrow \rho}\times \mathbb{Z}_2^{\tau\leftrightarrow -\bar\rho}\, .
\end{equation}
Here the group factor $SL(2)_\tau$ corresponds to the standard reparametrizations of the torus, acting as
\begin{equation}
\tau\rightarrow{a\tau+b\over c\tau+d}\, , \quad \textrm{with}\ \ \  ad-bc=1\,,
\end{equation}
whereas $SL(2)_\rho$ contains the shift in the $b$-field, $\rho\rightarrow \rho+c$, as well
as the  T-duality transformation (radius inversion), $\rho\rightarrow -1/\rho$, the full action being
\begin{equation}
\rho\rightarrow{a'\rho+b'\over c'\rho+d'}\, ,\quad \textrm{with} \ \ \ a'd'-b'c'=1\,.
\end{equation}
The embedding of $SL(2)_\tau\times SL(2)_\rho$ in $O(2,2)$ is then provided by the following identification with the matrices $A,B,C,D$ in eq. (\ref{SOdmatrix}):
\begin{equation}\label{embedding}
A=a'\begin{pmatrix} a & b \\ c & d \end{pmatrix}\, ,\quad B=b'\begin{pmatrix} -b & a \\ -d & c \end{pmatrix}
\, ,\quad C=c'\begin{pmatrix} -c & -d \\ a & b \end{pmatrix}\, ,\quad D=d'\begin{pmatrix} d& -c \\ -b & a \end{pmatrix}\,.
\end{equation}


\subsection{Symmetric $\mathbb{Z}_4$ orbifold: geometric T-fold  with one elliptic monodromy}

We first consider the case of a symmetric $\mathbb{Z}_4$-orbifold, where the  shift in the base coordinate $\mathbb{X}\rightarrow \mathbb{X} +2\pi$ is accompanied by the following symmetric $\mathbb{Z}_4$ rotation in the fiber:
\begin{equation}
\begin{split}
{\cal M}^{(1)}:
\end{split}
\quad
\begin{split}
X_L^{1\prime} &= -X_L^2 \; ,\\
X_L^{2\prime} &= X_L^{1}  \; ,
\end{split}
\qquad
\begin{split}
X_R^{1\prime} &=  - X_R^{2}\; ,\\
X_R^{2\prime} &=  X_R^{1}\, .
\end{split}
\end{equation}
This is indeed a left-right symmetric $\mathbb{Z}_4$ rotation, acting on the complex coordinates $Z_{L,R}=X^1_{L,R}+iX^2_{L,R}$ as
\begin{equation}
\begin{split}
{\cal M}^{(1)}:
\end{split}
\begin{split}
\quad Z_L^{\prime} &= e^{-{i\pi\over 2}}Z_L \; ,\\
 Z_R^{\prime} &=  e^{-{i\pi\over 2}}Z_R\, .
 \end{split}
 \label{monodLRc1}
\end{equation}
In the base $(X^I,\tilde{X}^I)$ of the fiber coordinates and their duals, $X^{1,2}={X^{1,2}_L+ X^{1,2}_R}$ and $\tilde X^{1,2}={X^{1,2}_L-X^{1,2}_R}$, the above $\mathbb{Z}_4$ transformation takes the form
\begin{equation}
\begin{split}
{\cal M}^{(1)}:
\end{split}
\begin{split}
\quad X_L^{1\prime} &= -X_L^2 \; ,\nonumber\\
X_L^{2\prime} &= X_L^{1}  \; ,\nonumber\\
\end{split}
\qquad
\begin{split}
X_R^{1\prime} &=  - X_R^{2}\; ,\nonumber\\
 X_R^{2\prime} &=  X_R^{1}\, ,
\end{split}  \label{monod1}
\end{equation}
and can be seen  to correspond to the following discrete $O(2,2)$ transformation:
\be
{\cal M}^{(1)}=
\begin{pmatrix}0&-1&0&0\\ 1&0&0&0\\0&0&0&-1\\0&0&1&0 \end{pmatrix}\,  .
\ee
Using the explicit embedding of $SL(2)_\tau\times SL(2)_\rho$ inside $O(2,2)$,  given in eq. (\ref{embedding}), one finds  that this transformation simply corresponds to an inversion of the complex structure:
\be\label{tautrans}
{\cal M}^{(1)}:\quad\tau'=\tau({\mathbb{X}+2\pi})=-{1\over \tau(\mathbb{X})}\, , \quad\rho'=\rho(\mathbb{X}+2\pi)=\rho(\mathbb{X})\, .
\ee
The corresponding geometric background can then be constructed  as a fibered torus with the following non-constant complex structure \cite{Hull:2005hk}:
\begin{align}
\tau(\mathbb{X})&={\tau_0\cos (f\mathbb{X})+\sin (f\mathbb{X})\over\cos (f\mathbb{X})-\tau_0\sin (f\mathbb{X})}\, ,\quad f\in{1\over 4}+{\mathbb Z}\, ,\nonumber\\
\rho(\mathbb{X})&=\rho_0=\textrm{const}\, .
\end{align}
Here $\tau_0$ and $\rho_0$ are arbitrary parameters (moduli) of the  background.
The background possesses geometric $\omega,\tilde\omega$-fluxes, which can be readily identified with $f$. At the fixed point of the  transformation $\mathbb{X}\rightarrow \mathbb{X}+2\pi$, $\tau_0=i$, the geometric background reduces precisely to the symmetric orbifold of Section \ref{momShift}. Note that the monodromy acts as an order two transformation on the background parameter $\tau(\mathbb{X})$, whereas its (symmetric) action on the coordinates $Z_L$ and $Z_R$ is of order
four.


\subsection{Asymmetric $\mathbb{Z}_4$ orbifold: non-geometric T-dual  T-fold with one elliptic monodromy}

We now start from the geometric background of the previous section and  perform a T-duality transformation along the $X^2$ direction of the fiber. This leads to a background, where the  shift in $\mathbb{X}$ is accompanied by the following
left-right asymmetric $\mathbb{Z}_4$ rotation, now acting on the complex coordinates $Z_{L,R}$ as
\begin{equation}
\begin{split}
{\cal M}^{(2)}:
\end{split}
\quad
\begin{split}
Z_L^{\prime} &= e^{-{i\pi\over 2}}Z_L \; ,\\
 Z_R^{\prime} &=  e^{{i\pi\over 2}}Z_R\, ,
  \end{split}\label{monodLRc2}
\end{equation}
which corresponds to the following $O(2,2)$ transformation:
\be
{\cal M}^{(2)}=
\begin{pmatrix}0&0&0&-1\\ 0&0&1&0\\0&-1&0&0\\1&0&0&0 \end{pmatrix}\,  .
\ee
As expected by T-duality, using  eq. (\ref{embedding}), it is straightforward to see that the action on the $T^2$-background now involves the inversion of the K\"ahler parameter, while preserving the complex structure unchanged
\be\label{rhotrans}
{\cal M}^{(2)}:\quad\tau'=\tau(\mathbb{X}+2\pi)=\tau(\mathbb{X})\, , \quad\rho'=\rho(\mathbb{X}+2\pi)=-{1\over \rho(\mathbb{X})}\, .
\ee
The corresponding non-geometric background is characterized by the following non-constant K\"ahler parameter:
\begin{eqnarray}
\tau(\mathbb{X})&=&\tau_0=\textrm{const}\, ,\nonumber\\
\rho(\mathbb{X})&=&{\rho_0\cos (g\mathbb{X})+\sin (g\mathbb{X})\over\cos (g\mathbb{X})-\rho_0\sin (g\mathbb{X})}\, ,\quad g\in{1\over 4}+{\mathbb Z}\, .
\end{eqnarray}
The background is a non-geometric T-fold and it possesses  both $H$-flux and non-geometric $Q$-flux. At the level of the above construction, $g$ should be thought of as a flux parameter (analogous to the eigenvalue of the flux matrix $F$), generating both the geometric $H$- and the non-geometric $Q$- flux.  The very fact that there exists one flux parameter $g$ generating both $H$ and $Q$, reflects the structure of the corresponding gauge algebra. At the fixed point of the  transformation $\mathbb{X}\rightarrow \mathbb{X}+2\pi$, $\rho_0=i$, the non-geometric background reduces to the asymmetric $\mathbb{Z}_4$ orbifold of Section \ref{SecTdualityFiber}.


\subsection{Truly asymmetric $\mathbb{Z}_2$ orbifold: non-geometric  T-fold with two elliptic monodromies}

We are now ready to describe an asymmetric orbifold, where the rotation acts only on the left-moving coordinates. As was already discussed in the previous sections, the resulting T-fold will be ``truly'' non-geometric, in the sence that it cannot be T-dualized to a geometric space\footnote{This background was already mentioned
in   \cite{Schulgin:2008fv}.    Genuinely non-geometric backgrounds and their relation to double geometry will be further discussed in \cite{HHLZ}.}. Specifically, let us consider  a background, where the  shift in $\mathbb{X}$
induces the following asymmetric $\mathbb{Z}_2$-rotation
on the complex coordinates $Z_{L,R}$
\begin{equation}
\begin{split}
{\cal M}^{(3)}:
\end{split}
\quad
\begin{split}
Z_L^{\prime} &= -Z_L \; ,\\
 Z_R^{\prime} &=  Z_R\, .
 \end{split} \label{monodLRc3}
\end{equation}
From the point of view of the $(X^I,\tilde{X}^I)$ basis, the orbifold action  mixes coordinates and dual coordinates as:
\begin{equation}
\begin{split}
{\cal M}^{(3)}:
\end{split}
\quad
\begin{split}
 X^{1\prime} &= -\tilde X^1 \; ,\\
X^{2\prime} &= -\tilde X^{2}  \; ,\\
\end{split}
\qquad
\begin{split}
\tilde X^{1\prime} &=   -X^{1}\; ,\\
\tilde X^{2\prime} &= -X^{2}\, ,
 \end{split} \label{monod3}
\end{equation}
and can be seen to correspond to the following $O(2,2)$ transformation:
\be\label{sodd}
{\cal M}^{(3)}=
\begin{pmatrix}0&0&-1&0\\ 0&0&0&-1\\-1&0&0&0\\0&-1&0&0 \end{pmatrix}\,  .
\ee
Upon using eq. (\ref{embedding}), one recognizes  that this transformation simultaneously induces
the inversion both of the K\"ahler parameter as well as of the complex structure:
\be\label{taurhotrans}
{\cal M}^{(3)}:\quad \tau'=\tau(\mathbb{X}+2\pi)=-{1\over \tau(\mathbb{X})}\, , \quad\rho'=\rho(\mathbb{X}+2\pi)=-{1\over \rho(\mathbb{X})}\, .
\ee
The corresponding background can be constructed as a double elliptic fibration with the following complex structure and K\"ahler parameters:
\begin{eqnarray}\label{functions}
\tau(\mathbb{X})&=&{\tau_0\cos (f\mathbb{X})+\sin (f\mathbb{X})\over\cos (f\mathbb{X})-\tau_0\sin (f\mathbb{X})}\, ,\quad f\in{1\over 4}+{\mathbb Z}\, ,\nonumber\\
\rho(\mathbb{X})&=&{\rho_0\cos (g\mathbb{X})+\sin (g\mathbb{X})\over\cos (g\mathbb{X})-\rho_0\sin (g\mathbb{X})}\, ,\quad g\in{1\over 4}+{\mathbb Z}\, .
\end{eqnarray}
As we have already mentioned, this  background is not T-dual to a geometric one. From the structure of the algebra in eq. \eqref{AlgMomentumShift}, it follows that the background contains $\omega,H$ and $Q$ flux. In the above construction, it is straightforward to identify $\omega, \tilde\omega$ with the $f$-parameter, whereas $H, Q$ are generated by $g$, consistently with the examples of the previous subsections. At the fixed point of the  transformation $\mathbb{X}\rightarrow \mathbb{X}+2\pi$, $\tau_0=\rho_0=i$, the T-fold reduces to the corresponding asymmetric $\mathbb{Z}_2$ orbifold.


\subsection{Truly asymmetric $\mathbb{Z}_4$ orbifold: non-geometric  T-fold with two elliptic monodromies}

We now consider the ``truly'' asymmetric $\mathbb{Z}_4$-orbifold with pure momentum shift, based on the freely-acting asymmetric orbifold CFT discussed in \cite{Condeescu:2012sp}, and constructed explicitly in Section \ref{SecExtremeAsymZ4}.  The  shift in $\mathbb{X}$
now acts as the following ``extremely'' asymmetric $\mathbb{Z}_4$ rotation
on the left-moving complex coordinates of the fiber:
\begin{equation}
\begin{split}
{\cal M}^{(4)}:
\end{split}
\quad
\begin{split}
 Z_L^{\prime} &= e^{-{i\pi\over 2}}Z_L \; ,\\
 Z_R^{\prime} &=  Z_R\, .
 \end{split}
\end{equation}
The corresponding $O(2,2)$ transformation now reads:
\be
{\cal M}^{(4)}=
{1\over 2}\begin{pmatrix}1&1&-1&1\\ -1&1&-1&-1\\-1&1&1&1\\-1&-1&-1&1 \end{pmatrix}\,  .
\ee
Note that $({\cal M}^{(4)})^4=I$ and also that the square of this matrix agrees with
the $O(2,2)$ matrix in eq. (\ref{sodd}) of the previous example:  $({\cal M}^{(4)})^2={\cal M}^{(3)}$, as expected by the fact that $\mathbb{Z}_2\subset\mathbb{Z}_4$. Furthermore, it should be noted that, although the entries of this matrix are no longer integer-, but half-integer- valued,  $\mathcal{M}^{(4)}$ is an allowed $O(2,2)$ transformation, satisfying the constraints in eq. (\ref{constraints}) and is a symmetry of the underlying CFT.

In order to determine how this $O(2,2)$ transformation acts on the background fields $\tau$ and $\rho$,
we again use the embedding eq. (\ref{embedding}) and obtain:
\begin{equation}\label{newtrans}
SL(2)_\tau:\quad \begin{pmatrix} a & b \\ c & d \end{pmatrix}={1\over\sqrt2 }\begin{pmatrix} 1 & 1 \\ -1 & 1 \end{pmatrix}\, ,\quad
SL(2)_\rho:\quad \begin{pmatrix} a' & b' \\ c' & d' \end{pmatrix}={1\over\sqrt2 }\begin{pmatrix} 1 & 1 \\ -1 & 1 \end{pmatrix}\, .
\end{equation}
The above transformations are not elements of $SL(2,\mathbb{Z})$ and their action
on the K\"ahler parameter as well as on the complex structure becomes:
\be\label{taurhotransa}
\tau'=\tau(\mathbb{X}+2\pi)={1+\tau(\mathbb{X})\over 1-\tau(\mathbb{X})}\, , \quad\rho'=\rho(\mathbb{X}+2\pi)={1+\rho(\mathbb{X})\over 1-\rho(\mathbb{X})}\, .
\ee
The corresponding T-fold background is again a double elliptic fibration with the following complex structure and K\"ahler parameters:
\begin{equation}\label{finalfunction}
\begin{split}
\tau(\mathbb{X})&={\tau_0\cos (f\mathbb{X})+\sin (f\mathbb{X})\over\cos (f\mathbb{X})-\tau_0\sin (f\mathbb{X})}\, ,\quad f\in{1\over 8}+{\mathbb Z} \\
\rho(\mathbb{X})&={\rho_0\cos (g\mathbb{X})+\sin (g\mathbb{X})\over\cos (g\mathbb{X})-\rho_0\sin (g\mathbb{X})}\, ,\quad g\in{1\over 8}+{\mathbb Z}\, .
\end{split}
\end{equation}
Similarly to the previous case, this  background is not T-dual to a geometric one. Note that the two functions $\tau(\mathbb{X})$ and $\rho(\mathbb{X})$ in eq. (\ref{finalfunction})
have similar form to those in eq. (\ref{functions}),  the difference being that the $f$- and $g$- fluxes are now quantized in units of $1/8$ and the transformation in eq. (\ref{finalfunction}) is of order four, in contrast to the order two transformation of eq. (\ref{functions}). Again, the interpretation of the flux parameters $f,g$ is similar to that of the previous subsection, with $f$ being identified with the geometric $\omega,\tilde\omega$ fluxes, whereas $g$ generates the geometric $H$- and non-geometric $Q$- fluxes. At the fixed point of the order four transformation  $\mathbb{X}\rightarrow \mathbb{X}+2\pi$, $\tau_0=\rho_0=i$, the T-fold reduces to the asymmetric $\mathbb{Z}_4$ orbifold of Section \ref{SecExtremeAsymZ4}.

From the effective supergravity point of view, the explicit expressions in eq. \eqref{finalfunction} can be cast in more intuitive form which reflects the asymmetric Scherk-Schwarz mechanism, by performing  holomorphic field redefinitions on the chiral scalar fields. To this end, consider the following field redefinition on the field $\tau$ (and similarly for $\rho$):
\begin{equation}
\tilde\tau={\tau-i\over \tau+i}\, ,
\end{equation}
such that at the fixed point $\tilde\tau$ vanishes: $\tilde\tau(\tau=i)=0$.
On the redefined field $\tilde\tau$, it follows that the $SL(2)$ transformation in eq. (\ref{taurhotransa}) has a simple action as a rotation by $\pi/2$:
\begin{equation}
\tilde\tau'=\tilde\tau(\mathbb{X}+2\pi)=i\tilde\tau(\mathbb{X})\, .
\end{equation}
Explicitly, in terms of the base coordinate:
\begin{equation}
\tilde\tau(\mathbb{X})=\frac{\tau_0-i}{\tau_0+i}\,\exp(2if\mathbb{X})\, .
\end{equation}
It should be noted that in the T-dual case (with respect to the base) of a winding shift, in which case the background contains $R$-flux, it is not possible to provide local expressions for $\tau$ and $\rho$, the monodromies would depend on shifts along the T-dual base coordinate $\tilde{\mathbb{X}}$.

Before closing this subsection, it is interesting to remark on deformation of the $\sigma$-model action away from the conformal point. Expanding eq. \eqref{finalfunction} around $\tau_0=\rho_0=i$, we may treat this non-geometric background using conformal perturbation theory, by representing the $\sigma$-model perturbation away from the orbifold point as an infinite series of irrelevant operators:
\begin{equation}
	\begin{split}
	S = S_{\textrm{orb}} + \sum_{\ell=1}^{\infty} \int d^2\sigma\, \bigr[(\tau_0-i)^\ell \sin^{\ell-1}(f\mathbb{X})\ e^{i(1+\ell)f\mathbb{X}}\ \partial Z \bar\partial\bar Z + \textrm{c.c.}\bigr] \\
	+ \sum_{\ell=1}^{\infty} \int d^2\sigma\, \bigr[(\rho_0-i)^\ell \sin^{\ell-1}(g\mathbb{X})\ e^{i(1+\ell)g\mathbb{X}}\ \partial Z \bar\partial Z + \textrm{c.c.}\bigr]
	\end{split}\,.
\end{equation}
The validity of the approach relies on the fact that the operators appearing in the above perturbation carry well-defined left- and right- moving conformal weights, being exponential functions of the free-boson $\mathbb{X}$. Note that in the case of the geometric orbifolds studied in the previous subsections, the corresponding deformation of the $\sigma$-model action involves only the complex structure $\tau_0$, since the K\"ahler (volume) modulus  still corresponds to a flat direction at the minimum (orbifold point) of the effective scalar potential.

The background we constructed in this section corresponds to the case of a pure momentum shift, which is clearly visible at the orbifold point and this is reflected by the fact that the fluxes $f,g$ only couple to the base coordinate $\mathbb{X}$, but not its dual. The T-dual version, with respect to the base, corresponding to the winding shift can be easily obtained by replacing $\mathbb{X}\rightarrow\tilde{\mathbb{X}}$. Of course, in this case the interpretation of the fluxes changes, according to the map in  eq. \eqref{mappingfluxes}. The more interesting case of simultaneous momentum and winding shift (\emph{e.g.} with a single $\mathbb{Z}_4$-orbifold) is very similar to the $\mathbb{Z}_4\times\mathbb{Z}_2$ background we discuss in the next section, the main difference (at the level of this discussion) being in the quantization of the fluxes.


\subsection{Asymmetric $\mathbb{Z}_4\times\mathbb{Z}_2$ orbifold: irreducible $R$-background with two elliptic monodromies}\label{Z4Z2backgr}

Finally, it is interesting to consider the asymmetric $\mathbb{Z}_4\times\mathbb{Z}_2$ orbifold, defined in Section \ref{Z2Z4orbCFT}. Since this background contains $R$-flux, one expects the description to be non-local and, hence, cannot be described as a T-fold. As before, we will focus our attention on one of the two $T^2$-tori inside the $T^4$ fiber. The shift in $\mathbb{X}$ is accompanied a purely left-moving $\mathbb{Z}_4$ rotation of the fiber, whereas the shift in $\tilde{\mathbb{X}}$ is accompanied by a $\mathbb{Z}_2$ rotation of the right-movers:
\begin{equation}
\begin{split}
{\cal M}^{(4)}:
\end{split}
\quad
\begin{split}
 Z_L^{\prime} &= e^{-{i\pi\over 2}}Z_L \; ,\\
 Z_R^{\prime} &=  Z_R\, .
 \end{split} \qquad
\begin{split}
{\cal M}^{(2)}:
\end{split}
\quad
\begin{split}
 Z_L^{\prime} &=  Z_L \; ,\\
 Z_R^{\prime} &= - Z_R\, .
 \end{split}\, ,
\end{equation}
corresponding to the following $O(2,2)$ transformation:
\be
{\cal M}^{(4)}=
{1\over 2}\begin{pmatrix}1&1&-1&1\\ -1&1&-1&-1\\-1&1&1&1\\-1&-1&-1&1 \end{pmatrix} \qquad,\qquad
{\cal M}^{(2)}=
\begin{pmatrix}0&0&1&0\\ 0&0&0&1\\1&0&0&0\\0&1&0&0 \end{pmatrix}\,  .
\ee
Combining the results from the previous subsections, we may now write down the action on the K\"ahler and complex structure moduli:
\be
\begin{split}
\tau(\mathbb{X}+2\pi,\tilde{\mathbb{X}})={1+\tau(\mathbb{X},\tilde{\mathbb{X}})\over 1-\tau(\mathbb{X},\tilde{\mathbb{X}})}\ &, \quad \rho(\mathbb{X}+2\pi,\tilde{\mathbb{X}})={1+\rho(\mathbb{X},\tilde{\mathbb{X}})\over 1-\rho(\mathbb{X},\tilde{\mathbb{X}})} \\
 \tau(\mathbb{X},\tilde{\mathbb{X}}+2\pi)=-{1\over \tau(\mathbb{X},\tilde{\mathbb{X}})}\ &, \quad \rho(\mathbb{X},\tilde{\mathbb{X}}+2\pi)=-{1\over \rho(\mathbb{X},\tilde{\mathbb{X}})}
 \end{split}\,.
\ee
The corresponding background is again a double elliptic fibration in $\mathbb{X},\tilde{\mathbb{X}}$, with the following complex structure and K\"ahler parameters:
\begin{equation}
\begin{split}
\tau(\mathbb{X},\tilde{\mathbb{X}})&={\tau_0\cos (f_4\mathbb{X}+f_2\tilde{\mathbb{X}})+\sin (f_4\mathbb{X}+f_2\tilde{\mathbb{X}})\over\cos (f_4\mathbb{X}+f_2\tilde{\mathbb{X}})-\tau_0\sin (f_4\mathbb{X}+f_2\tilde{\mathbb{X}})}\ ,\quad f_4,g_4\in{1\over 8}+{\mathbb Z}\\
\rho(\mathbb{X},\tilde{\mathbb{X}})&={\rho_0\cos (g_4\mathbb{X}+g_2\tilde{\mathbb{X}})+\sin (g_4\mathbb{X}+g_2\tilde{\mathbb{X}})\over\cos (g_4\mathbb{X}+g_2\tilde{\mathbb{X}})-\rho_0\sin (g_4\mathbb{X}+g_2\tilde{\mathbb{X}})}\ , \quad f_2, g_2\in{1\over 4}+{\mathbb Z}
\end{split}\,.
\end{equation}
Notice that the expressions for the K\"ahler and complex structure moduli are inherently non-local, as expected from the fact that the algebra of gaugings in eq. \eqref{Z2Z4algebra} contains $R$-flux. The parameters $f_2,g_2,f_4,g_4$  generate the  following irreducible combinations of fluxes:
\begin{table}[h]
	\centering
	\begin{tabular}{|c | c |}\hline
	Parameter & ~Fluxes~ \\ [0.5ex] \hline \hline
	$f_4$ & $\omega , \tilde\omega$\\   [0.6ex]
	$f_2$ & $Q , \tilde{Q}$\\   [0.6ex]
	$g_4$ & $H , Q$\\  [0.6ex]
	$g_2$ & $\tilde\omega , R$ \\ [0.6ex] \hline 
	\end{tabular}\label{table1}
\end{table}

The above identification is consistent with the fact that the $H$-flux originates from the K\"ahler parameter $\rho$ and with the general mapping of the fluxes under a T-duality in the base, as in eq. \eqref{mappingfluxes}.

At the fixed point of the order four- and order two- transformations  $\mathbb{X}\rightarrow \mathbb{X}+2\pi$, $\tilde{\mathbb{X}}\rightarrow \tilde{\mathbb{X}}+2\pi$, $\tau_0=\rho_0=i$, and the background reduces to the asymmetric $\mathbb{Z}_4$ orbifold of Section \ref{Z2Z4orbCFT}. As before, one may perform the following field redefinition on $\tau$ (and similarly for $\rho$):
\begin{equation}
\tilde\tau={\tau-i\over \tau+i}\, ,
\end{equation}
in which case one obtains the doubled space analogue of the twisted reduction:
\begin{equation}
\tilde\tau(\mathbb{X})=\frac{\tau_0-i}{\tau_0+i}\,\exp\bigr[2i(f_4\mathbb{X}+f_2\tilde{\mathbb{X}})\bigr]\, ,
\end{equation}
in agreement with eq. \eqref{momwindreduction}.

Finally, a deformation away from the orbifold point can be formulated at the level of the string $\sigma$-model in terms of conformal perturbation theory. Namely, one may consider perturbing the $\sigma$-model action at the conformal orbifold point by inserting the following infinite series of irrelevant operators:
\begin{equation}
	\begin{split}
	S = S_{\textrm{orb}} + \sum_{\ell=1}^{\infty} \int d^2\sigma\, \bigr[(\tau_0-i)^\ell \sin^{\ell-1}(f_4\mathbb{X}+f_2\tilde{\mathbb{X}}_2)\ e^{i(1+\ell)(f_4\mathbb{X}+f_2\tilde{\mathbb{X}}_2)}\ \partial Z \bar\partial\bar Z + \textrm{c.c.}\bigr] \\
	+\sum_{\ell=1}^{\infty} \int d^2\sigma\, \bigr[(\rho_0-i)^\ell \sin^{\ell-1}(g_4\mathbb{X}+g_2\tilde{\mathbb{X}}_2)\ e^{i(1+\ell)(g_4\mathbb{X}+g_2\tilde{\mathbb{X}}_2)}\ \partial Z \bar\partial Z + \textrm{c.c.}\bigr]
	\end{split}\,.
\end{equation}
Here, the circle coordinate $\mathbb{X}$ and its dual $\tilde{\mathbb{X}}$ should be thought of as linear combinations of $\mathbb{X}_{L}$ and $\mathbb{X}_R$. Since the perturbation depends on the left- and right- moving circle coordinates only through exponential functions, the corresponding operators $\mathcal{O}_{q_L,q_R}\sim e^{iq_L \mathbb{X}_L+iq_R\mathbb{X}_R}$ carry well-defined conformal weight  and the deformation can be consistently defined at the string level, for any circle radius $R$. It would be interesting to obtain an analogous formulation of this deformation in terms of a double $\sigma$-model, along the lines of \cite{Hull:2006qs,Copland:2011wx,Nibbelink:2012jb}.

\section{Summary}

In this paper we have explicitly constructed freely acting orbifold CFT's that correspond to non-geometric string backgrounds with $Q$- and $R$-fluxes.
The fluxes are identified from the CFT operator algebras and by comparison with the associated flux algebras in the effective gauged supergravity theory.
In particular, we provide for the first time explicit backgrounds with simultaneous $Q$- and $R$-fluxes,  obtained by combining shifts and rotations acting
asymmetrically both in the base as well as in the fiber directions. In this way we obtain generalized $Q$-flux (T-fold) and $R$-flux backgrounds that cannot be T-dualized to geometric ones. In the case of ``true'' $R$-backgrounds, the corresponding fields of the fiber
simultaneously depend both on the momentum as well as on the winding (dual) coordinate of the base. We expect these inherently non-local backgrounds to admit a natural description in terms of double field theory and it would be interesting for this connection to be made more precise \cite{HHLZ}.


\subsection*{Acknowledgements}

We are grateful to G. Dall'Agata, F. Ha\ss ler, O. Hohm, C. Hull, K. Siampos, N. Toumbas, A. Zein Assi and  B. Zwiebach for several fruitful discussions. C.C. is thankful to the Max-Planck-Institut f\"ur Physik in Munich for hospitality during the early stages of this work. I.F. would like to thank the CERN Theory Division, as well as the University of Rome ``Tor Vergata" for hospitality. I.F. and D.L. wish to thank the Laboratoire de Physique Th\'eorique de l'Ecole Normale Sup\'erieure for its hospitality during the last stages of this work. This work is supported by the ERC Advanced Grant ``Strings and Gravity" (Grant.No. 32004), by the DFG Transregional Collaborative Research Centre ``Dark Universe" (TRR 33), the DFG cluster of excellence ``Origin and Structure of the Universe" and by the CNCS-UEFISCDI grant PD 103/2012.


\bigskip
\medskip

\bibliographystyle{unsrt}

\vfill\eject
\end{document}